\documentclass{article}

\usepackage{arxiv}

\usepackage[toc,page]{appendix}

\usepackage[utf8]{inputenc} % allow utf-8 input
\usepackage[T1]{fontenc}    % use 8-bit T1 fonts
\usepackage{hyperref}       % hyperlinks
\usepackage{url}            % simple URL typesetting
\usepackage{booktabs}       % professional-quality tables
\usepackage{amsfonts}       % blackboard math symbols
\usepackage{nicefrac}       % compact symbols for 1/2, etc.
\usepackage{microtype}      % microtypography
\usepackage{lipsum}

\usepackage{graphicx}
\usepackage{graphicx}
\usepackage{float}

\usepackage{algorithm}
\usepackage{algorithmic}
\usepackage{amsmath}
\usepackage{multicol}

\usepackage[utf8]{inputenc}
\usepackage{multicol}
\usepackage{subfig}

\title{An Epidemiological Modelling Approach for Covid19 via Data Assimilation}

\author{
 Philip Nadler\thanks{Corresponding Author} \\
  Data Science Institute\\
  Imperial College London\\
  London SW7 2AZ, U.K. \\
  \texttt{p.nadler@imperial.ac.uk} \\
   \And
 Shuo Wang \\
  Data Science Institute\\
  Imperial College London\\
  London SW7 2AZ, U.K. \\
  \texttt{shuo.wang@imperial.ac.uk} \\
     \And
 Rossella Arcucci \\
  Data Science Institute\\
  Imperial College London\\
  London SW7 2AZ, U.K. \\
  \texttt{r.arcucci@imperial.ac.uk} \\
      \And
 Xian Yang \\
  Data Science Institute\\
  Imperial College London\\
  London SW7 2AZ, U.K. \\
  \texttt{xian.yang08@imperial.ac.uk} \\
       \And
Yike Guo* \\
  Data Science Institute\\
  Imperial College London\\
  London SW7 2AZ, U.K. \\
  \texttt{y.guo@imperial.ac.uk} \\
  }
\begin{document}
\maketitle

\begin{abstract}
%\lipsum[1]
The global pandemic of the 2019-nCov requires the evaluation of policy interventions to mitigate future social and economic costs of quarantine measures worldwide. We propose an epidemiological model for forecasting and policy evaluation which incorporates new data in real-time through variational data assimilation. We analyze and discuss infection rates in the UK, US and Italy. We furthermore develop a custom compartmental SIR model fit to variables related to the available data of the pandemic, named SITR model, which allows for more granular inference on infection numbers. 
We compare and discuss model results which conducts updates as new observations become available. A hybrid data assimilation approach is applied to make results robust to initial conditions and measurement errors in the data. We use the model to conduct inference on infection numbers as well as parameters such as the disease transmissibility rate or the rate of recovery. The parameterisation of the model is parsimonious and extendable, allowing for the incorporation of additional data and parameters of interest. This allows for scalability and the extension of the model to other locations or the adaption of novel data sources.
\end{abstract}

% keywords can be removed
\keywords{Data Assimilation \and 2019-nCov \and Inference \and Bayesian Updating \and Compartmental Model}
\newpage
%\tableofcontents
\newpage

\section{Introduction}
The global outbreak of n-Cov2019  and the possibility of severe social and economic costs worldwide requires immediate action on suppresion measures. In order to evaluate the efficacy of past and future policy measures to fight and contain the spread of n-Cov2019, a robust and quantifiable analysis system is required. We propose a methodology for forecasting the spread of n-Cov2019 and show how to estimate latent infection rates, accounting for high uncertainty in observation and model specification, which is done by combining real-time Bayesian updating with epidemiological models. \newline
To show the generalisability of our updating approach we first embed a standard SIR model in our framework and then develop a custom compartmental SIR model which is fit to data related to the spread of the coronavirus worldwide which we name SITR. The SITR model adds an additional compartment for patients under treatment $T$ and allows for more granular inference on the underlying dynamics of the epidemic, separating confirmed cases under treatment with latent unconfirmed cases of Covid19.
The models are embedded in a data assimilation framework, a form of recursive Bayesian estimation \cite{asch_2016}, which conducts model updates when new observations become available. The assimilation scheme lends itself naturally to the problem because the procedure allows the model to dynamically adjust infection rates in real time, while taking into account the uncertainty in the data via the specification of covariance matrices.\newline
The uncertainty quantification and choice of covariance matrices is being analyzed using a hybrid data assimilation approach, which is applied to make results robust to initial conditions. We use the model to infer the amount of infected people and both, the disease transmissibility rate, as well as the rate of recovery. The time-varying parameter structure of the model allows for the incorporation and analysis of policy action, such as if the shutdown of transportation or closure of schools affect transmissibility.\newline
In line with other researchers, our model estimates indicate that the number of infected people is a number of magnitudes higher than the actual reported number of confirmed reported infections. We also find that compared to static models, updating the parameters in a dynamic fashion leads to an upward correction of the true number of infected people as well as reducing forecasting errors.\newline
We estimate both short term and long term dynamics in Italy, the United States and the United Kingdom, finding a peak of infections in the middle of March and a flattened but sustained number of infected cases in the United States and the United Kingdom. We furthermore analyze the transmissibility rate and find that they decreased after imposed initial lockdown measures, but increased again after a loosening of restrictions end of May. The rest of the paper is structured as follows: Section two discusses related work. Section three and four introduce the dynamic model as well as the SITR compartmental model. Section five discusses how the uncertainty in the data is incorporated in the model. Section six states discusses empirical results and section seven concludes.
 %The model ingests new observations based on confirmed infection cases and estimates unobservable variables such as the total amount of infected people or the transmissibility of the virus. The time-varying structure of the model allows for the incorporation and analysis of policy action, such as if the shutdown of transportation or closure of schools affect transmissibility. This allows for policy recommendations and predicting the effects of possible future government interventions. The model can be extended for global analysis and initial model results can be enhanced with more granular data.

\section{Related Work}
The spread of the novel type of respiratory virus as well as the dramatic economic consequences trying to contain it has led to a rapid engagement of the scientific community, with many different areas of research being explored.\newline
Using compartmental models in epidemiology, authors such as \cite{imai2020estimating}, \cite{li2020early} and \cite{wu2020nowcasting} have done the first studies on the size of the outbreak in China. They applied standard SIR models with static parameters to estimate the basic reproductive number and analyze the exponential growth of the virus in Wuhan. The work of \cite{wu2020nowcasting} in particular combines standard SEIR models with travel data obtained from Tencent and found that epidemic dynamics show exponential patterns in multiple major cities with a lag behind the Wuhan outbreak of about one to two weeks.\newline
First studies using data assimilation for epidemiological modelling have been conducted by other authors such as \cite{rhodes2009variational} and \cite{bettencourt2007towards}, which studied the techniques on different cases such as influenza using standard SIR models, although none has considered the issue of the robust covariance estimates as discussed in \cite{wang2013gsi} or \cite{bonavita2016evolution}.\newline 
Further studies such as \cite{bettencourt2008real} and \cite{cobb2014bayesian} study time varying parameters in more detail, although only for standard SIR models with no relationship to the current corona virus outbreak. We are the first to conduct a study of the current spread of 2019-nCoV using data assimilation. We furthermore contribute by providing a novel framework which enables the prior computation of covariance matrices, adding robustness to epidemiological assimilation models.
Although many compartmental models are available, we base our initial studies on SIR models, since it allows us to verify the dynamics of the assimilation scheme as a robust benchmark and compare it to extensions of the model
later on. We will specify the exact model choice and specification in the next section.

% PN: This will be a contribution to the current work, cite arxiv paper.

% We apply the methodology to a SIR model. The methods is general and it can be applied to other models such as SITR,... . In our case, we focussed on a standard SIR model because ...........

% Why do we not include S-E-IR, the E compartment for latent infection period?
% We want to keep the model at this point. Furthermore, the data "confirmed" cases is mostly for people who are tested and thus symptomatic already. Asymptomatic cases are rarely tested because of a general lack of testing equipment , especially in the early development of the virus.

% E can be extension in the future with more data and if testing practices change, at this point it would increase model complexity with little gained explainability and possibly bias other coefficient estimates.

\section{The Adaptive Epidemiological Model}

We introduce an adaptive epidemiological modelling framework which combines a SIR model whose model parameters are time-varying with data assimilation techniques.
%PN: update on SEIR model
We base our model on the most basic compartmental model, which is the SIR. Describing and implementing the assimilation scheme in the basic structure of a SIR model allows us to analyze its initial performance and derive additional modifications later on. 
Further complexities will be introduced when more granular data is available. The current confirmed cases are mostly symptomatic cases with many asymptomatic cases being unconfirmed due to limited testing capacities.\newline
For Covid19, a SEIR model adding an compartment for exposure is a possible candidate for further extensions. But given the lack of more granular data with all confirmed infection cases being symptomatic and no other data on exposure levels, adding additional parameters with insufficient data can lead to a bias of parameter estimates in the assimilation procedure, and also increases model complexity without adding understanding of the underlying mechanics of the Covid19 infections \cite{li1995global, seibert2019validation}. The lack of meaningful and accurate data which fits the model assumptions for more complex models strongly affects the performance of data assimilation and thus makes a sparse and parsimonious model preferable \cite{asch_2016}. 

\subsection{The Standard SIR Model}
We start our analysis with a standard SIR model \cite{anderson1991discussion}, which is a system of three interrelated non-linear ordinary differential equations without an explicit analytical solution.
The dynamics of the model are given by :
\begin{equation}\label{eq:SIR_S}
\frac{dS}{dt} = -\beta\frac{IS}{N}  
\end{equation}{}
\begin{equation}\label{eq:SIR_I}
\frac{dI}{dt} = \beta\frac{IS}{N}-\gamma I
\end{equation}{}
\begin{equation}\label{eq:SIR_R}
\frac{dR}{dt} =\gamma I
\end{equation}{}

Where $S$ denotes the susceptible population size, $I$ the infected people who are not isolated from the population and $R$ the recovered population. The total population is given by $N$. The parameters $\beta$ and $\gamma$ denote the transmission and recover rate of the virus infection. Note that for the outbreak, the susceptible number $S$ is observable, which we label as the population of the country under analysis. The recovered population $R$ denotes the population not infectious anymore and being removed from the population, which for the studied examples is the number of confirmed cases, since confirmed cases are hospitalized and isolated and not infecting the general population anymore.

%\subsection{The Data Assimilation Model}
\subsection{The Adaptive DA-SIR model}
Data Assimilation (DA) is a technique to incorporate observations into a theoretical model where uncertainty is quantified~\cite{asch_2016}. It allows for problems with uneven spatial and temporal data distribution and redundancy to be addressed such that models can ingest information. DA is a vital step in numerical modeling and has become a main component in the development and validation of mathematical models in meteorology, climatology, geophysics, geology and hydrology \cite{cuomo2017numerical}. Recently, DA is also applied to numerical simulations of geophysical applications, medicine, biological science and finance \cite{nadler2019scalable}. 
Data assimilation can be applied to a variety of problems where an uncertainty quantification has to be included~\cite{arcucci2017variational} or where latent parameters need to be computed taking into account new observations.\newline
The Adaptive DA-SIR model is a model which incorporates data assimilation with a compartmental SIR model. We use DA as an adaptive modelling approach which integrates new observations into our compartmental model to enhance the accuracy of forecasts as well as computing model parameters of interest, in our case $\beta$ and $\gamma$ in the SIR model.
%The general assimilation model is described by the following equation
%\begin{equation}\label{eq:1}
%    \textbf{x}_{k+1} = \boldsymbol{\mathcal{M}}_{k+1}\textbf{x}_{k}
%\end{equation}
%where $\textbf{x}_{k}$ and $\boldsymbol{\mathcal{M}_{k}}$ are the state vector and nonlinear model operator at timestep $k$ respectively. Furthermore, $\textbf{y}^{o}_{k}$ represents the vector of observations at timestep $k$ and $\boldsymbol{\mathcal{H}}_{k}$ is the nonlinear observation operator that maps observation to model space:
%\begin{equation}\label{eq:2}
%    \textbf{y}_{k}^{o} = \boldsymbol{\mathcal{H}}_{k}\textbf{x}_{k}
%^\end{equation}

The SIR model in equations \eqref{eq:SIR_S}-\eqref{eq:SIR_R} can be discretized with respect to the time variable, giving the following equations:

\begin{equation}\label{eq:SIR_St}
S_{t+1} = S_t -\beta\frac{I_tS_t}{N}  
\end{equation}{}
\begin{equation}\label{eq:SIR_It}
I_{t+1} = I_t+ \beta\frac{I_tS_t}{N}-\gamma I_t
\end{equation}{}
\begin{equation}\label{eq:SIR_Rt}
R_{t+1} =R_t +\gamma I_t
\end{equation}{}

For a given time step $t$ and assuming to have observations of the variable $R_t$ we denote here with $R^{obs}_t$, the DA problem consists in computing the minimum of the cost function

\begin{equation}\label{eq:3}
J(I) =  \sum^{t+\tau}_{i=t+1} ||R^{obs}_i-R^{pred}_i(I,\beta, \gamma)||_{\textbf{Q}^{-1}_t} + ||I-I^{pred}_{t}||_{\textbf{P}^{-1}_{t}}
%\Vert R-R_t \Vert_{\textbf{Q}^{-1}}+\sum_{t^\star=t}^{t+M}\Vert R-R^{obs}_{t^\star} \Vert_{\textbf{P}^{-1}}
%J(I^DA) = \operatorname*{argmin}_{I} \sum^{t+\tau}_{i=t+1} ||R^{obs}_t-H(i|I,\beta, \gamma)||_{\textbf{Q}^{-1}_t} + ||I-I_{t}||_{\textbf{P}^{-1}_{t}}
\end{equation}
and
%\begin{equation}\label{eq:DA_Rt}
%    R^{DA}_t = argmin_R J(R)
%\end{equation}
\begin{equation}\label{eq:DA_Rt}
I^{DA}_{t}  = \operatorname*{argmin}_{I} J(I)
     %\sum^{t+\tau}_{i=t+1} ||R^{obs}_t-H(i|I,\beta_{t-1})||_{\textbf{Q}^{-1}_t} + ||I-I_{t}||_{\textbf{P}^{-1}_{t}}
\end{equation}{}
where $R^{pred}$ is a predicted value generated by the SIR model, and where $\textbf{Q}$ and $\textbf{P}$ denote the the background and the observation covariance matrices, representing an estimation of the errors in the data. 
To estimate the parameter, we minimize
\begin{equation}
  %  \beta_{t}^{e}  = \operatorname*{argmin}_{\beta^{e}} =\sum^{t+\tau}_{i=t+1} ||\textbf{H}(i)-\hat{\textbf{H}}(i|I,\beta^{e}_{t-1})||_{\textbf{Q}^{-1}_t} %+L(\beta^{e})
     \beta_t, \gamma_t  = \operatorname*{argmin}_{\beta, \gamma } \sum^{t+\tau}_{i=t+1} ||R^{obs}_i-R^{pred}_i(I_t^{DA},\beta, \gamma)||_{\textbf{Q}^{-1}_t} 
\end{equation}{}

Data assimilation is very sensitive to initial conditions and the choices of the covariance matrices, since they quantify uncertainty and determine how much weight is assigned to new observations which are assimilated into the model. Thus their calibration needs to be properly chosen, which we outline in detail in section 5. % as we will show in the following sections. %The cost function is a form of Tikhonov regularisation \cite{dong_2015} and is defined as
%\begin{align}\label{eq:3}
 %   J(\textbf{x})&=\frac{1}{2}\left(\textbf{x}-\textbf{x}^{b}\right)^{T}\textbf{Q}^{-1}\left(\textbf{x}-\textbf{x}^{b}\right) \nonumber \\
  %  &+ \frac{1}{2}\left(\textbf{H}\textbf{x}-\textbf{y}^{o}\right)^{T}\textbf{P}^{-1}\left(\textbf{H}\textbf{x}-\textbf{y}^{o}\right)
%\end{align}

% \begin{equation}\label{eq:3}
% J(\textbf{x}) = \Vert \textbf{x}-\textbf{x}^{b} \Vert_{\textbf{Q}^{-1}}+\sum_{time\ window} \Vert \textbf{H}\textbf{x}-\textbf{y}^{o} \Vert_{\textbf{P}^{-1}}
% \end{equation}

%\begin{equation}\label{eq:3}
%J(\textbf{x}) = \Vert \textbf{x}-\textbf{x}^{b} \Vert_{\textbf{Q}^{-1}}+\sum_{{\tiny window}}\Vert \textbf{H}\textbf{x}-\textbf{y}^{o} \Vert_{\textbf{P}^{-1}}
%\end{equation}
%where $\textbf{H}$ is the linearized version of the nonlinear observation operator $\boldsymbol{\mathcal{H}}$ with the background state vector give by $\textbf{x}^{b}$. The window size determines the number of past observations for updating and depends on the assimilation approach \cite{lorenc2003potential}. We next show how epidemiological compartmental models can be embedded in this approach.

%\subsection{The Adaptive DA-SIR model}
%We incorporate the dynamics in a assimilation framework where the state vector is given by: $\textbf{x}^{T} = [S,I,R]$. 

The data we use representing $S_t$, $I_t$ and $R_t$ is given by the official government numbers and is available at \cite{JHUdata, CDCdata}.
%To embed the model in the DA scheme given by Eq.~\ref{eq:1} to Eq.~\ref{eq:2}, minimization of the cost function Eq.~\ref{eq:3} 
The solution of the DA problem in \eqref{eq:3} leads to a modified extended Kalman filtering algorithm where an SIR model is used to compute the forward steps, e.g. in the time window $[t,t+M]$. Where $I_t^{DA}$ are the values of $I_t$ computed after the assimilation of $R_t^{obs}$ as in Eq.~\ref{eq:DA_Rt}. %The observation operator is given by $\textbf{H} = [0,0,1]$, since for the current corona virus outbreak the number of infections is unobservable. Only a fraction of confirmed cases $R$ are reported, thus only accumulated confirmed cases are observed and assimilated as given by $\textbf{H}$ . \\
%\subsection{Comparing our model to static models}
%Values of $R_t^{DA}$ computed by \eqref{eq:DA_Rt} affect the values of $I_t$ in the SIR model (equations \eqref{eq:SIR_St}-\eqref{eq:SIR_Rt}). We denote with $I_t^{DA}$ the values of $I_t$ computed after the assimilation of $R_t^{obs}$.
To illustrate and put results into perspective, we compare results of our adaptive DA-SIR model with the common SIR model for an example case.
\begin{figure}[h]%[htpb!]
    \centering
    \includegraphics[width=0.65\textwidth]{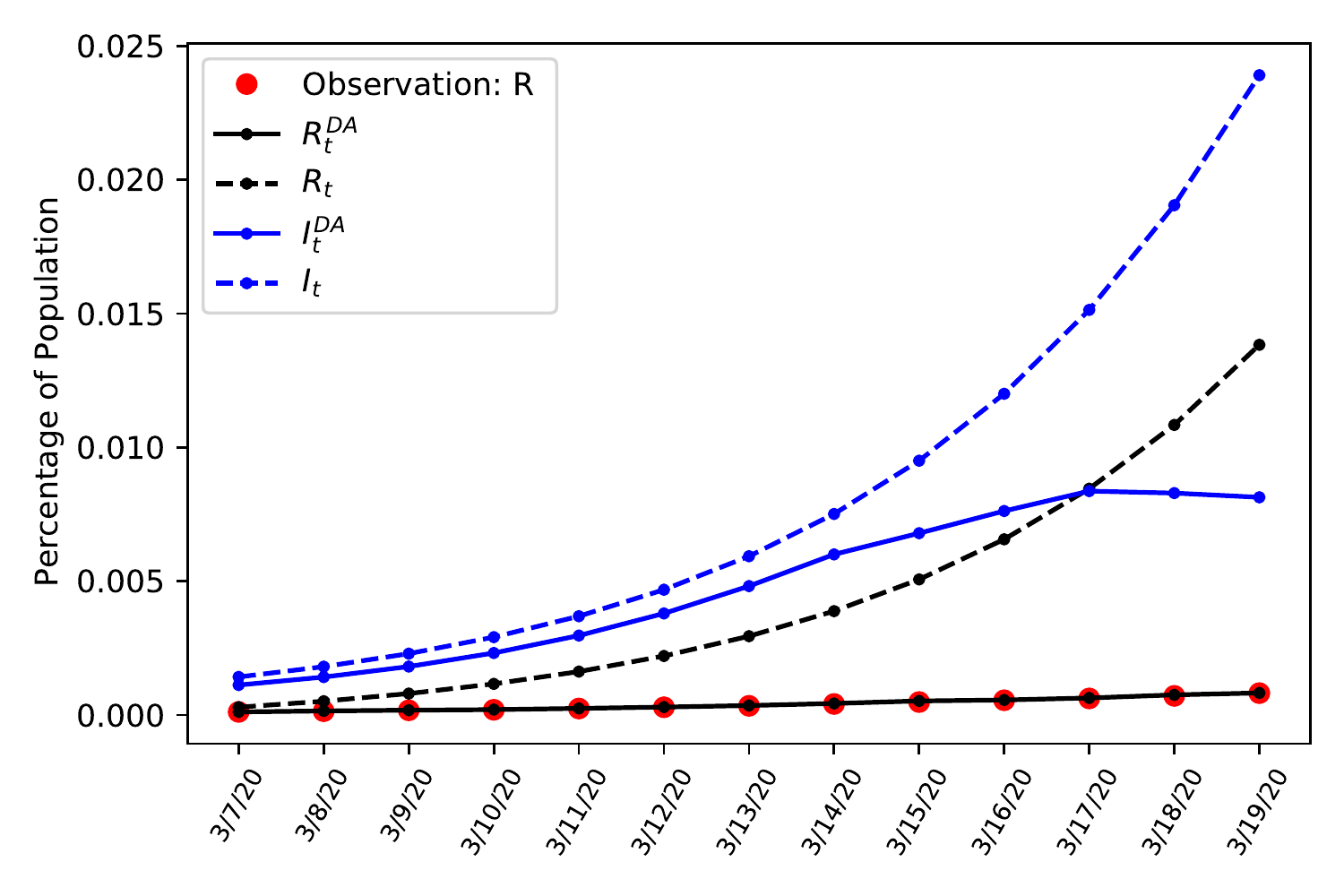}
    \caption{Comparing estimates during early stages of the outbreak of confirmed cases R and unobservable amount of infected people $I$ in Italy, using static and updating parameters.}
    \label{fig:staticvsdynamic}
\end{figure}{}
Both models use the same initial conditions given by the observed data. In Fig.~\ref{fig:staticvsdynamic} we compare model performance of the standard SIR model and show how assimilation of new observations generates updated model dynamics in the DA-SIR model that do differ from standard SIR model predictions by a wide margin, as is illustrated in the figure. Selected values of the graph are available in Table~\ref{tab:table_ODE} where we compare estimated confirmed cases and latent infection rates for both the static SIR, and dynamic DA-SIR model.

% \begin{table}[h]
% \centering
% % \label{features}
% \begin{tabular}{l|c|c|c|c|c|c|c}
% %\toprule
% \hline
% date &  01-15 &  01-19 &  01-23 &  01-27 &  01-31 &  02-03 &  02-07 \\
% \hline
% %\midrule
% $I_t^{DA}$  &      131 &      286 &     1182 &     4778 &    16130 &    30744 &    49801 \\
% $I_t$ &      166 &      342 &      893 &     2333 &     6093 &    12514 &    32656 \\ \hline
% $R_t^{DA}$  &       74 &      177 &      498 &     1338 &     3532 &     7279 &    19040 \\
% $R_t$ &       54 &      132 &      635 &     2700 &     7615 &    14491 &    26767 \\
% \hline
% %\bottomrule
% \end{tabular}

% \caption{Selected data points for predicted number of infected and treated patients for a dynamic model and a static ODE model}
% \label{tab:table_ODE}
% \end{table}{}

% \begin{table}[h]
% \centering
% % \label{features}
% \begin{tabular}{l|c|c|c|c|c|c|c}
% %\toprule
% \hline
% date &  03-07 &  03-09 &  03-11 &  03-12 &  03-14 &  03-16 &  18 \\
% \hline
% %\midrule
% $I_t^{DA}$  &      131 &      286 &     1182 &     4778 &    16130 &    30744 &    49801 \\
% $I_t$ &      166 &      342 &      893 &     2333 &     6093 &    12514 &    32656 \\ \hline
% $R_t^{DA}$  &       74 &      177 &      498 &     1338 &     3532 &     7279 &    19040 \\
% $R_t$ &       54 &      132 &      635 &     2700 &     7615 &    14491 &    26767 \\
% \hline
% %\bottomrule
% \end{tabular}

% \caption{Selected data points for predicted number of infected and treated patients for a dynamic model and a static ODE model}
% \label{tab:table_ODE}
% \end{table}{}

\begin{table}[h]
\centering
% \label{features}
\begin{tabular}{l|c|c|c|c|c|c|c}
%\toprule
\hline
date &    3/7/20 &     3/9/20 &    3/11/20 &    3/13/20 &    3/15/20 &    3/17/20 &     3/19/20 \\
\midrule
$I_t^{DA}$   &  67050 &  108540 &  178152 &  289418 &  422682 &  525755 &   534073 \\
$I_t$ &  85190 &  137428 &  221401 &  355915 &  570180 &  908422 &  1434817 \\ \hline
$R_t^{DA}$  &  17311 &   47899 &   97212 &  176569 &  303921 &  507379 &   830126 \\
$R_t$  &   6705 &   10582 &   14595 &   21030 &   30945 &   37853 &    49265 \\
\hline
%\bottomrule
\end{tabular}

\caption{Selected data points for predicted number of infected and treated patients for a dynamic model and a static ODE model}
\label{tab:table_ODE}
\end{table}{}

The dynamic model given by the solid lines fit the observed values of confirmed cases $R_t$ and interpolates the number of infected people $I_t$. The dashed lines represent the standard SIR model and show how not updating the model from the initial conditions leads to overestimation of infectious cases if interpolating according to simple exponential SIR dynamics, with a inferior model fit when comparing the observed confirmed cases denoted in red.
This illustrates how without updating the parameters the number of infected people is overestimated and the assimilation of new observations helps to adjust the trajectory of likely infections in the future. Having shown the large difference between static and dynamic SIR models we next introduce a further refined extension of the dynamic assimilation model.

\section{The Extended Epidemiological Assimilation Scheme}
%\section{The Extended Epidemiological Adaptive Assimilation Scheme}
\subsection{The SITR model}
Having illustrated the benefits of embedding the SIR model in a DA framework, we aim to further exploit the available data to do more fine tuned inference.
In the previous case of the simple SIR model, both recovered and isolated patients were categorized as $R$.
We revise the SIR model by introducing an intermediate compartment $T$. Here, $T$ represents the number of people being treated, given by the difference between accumulated confirmed cases and recovered or deceased patients $R$. Instead of just observing one variable, the number of confirmed cases, we are now observing two variables: the currently confirmed cases being treated $T$ and removed infectious population due to recovery or being deceased $R$. 
%The introduced additional data available from official sources and add an additional variable for treated people.
The model is given by
\begin{equation}\label{eq:SITR1}
\frac{dS}{dt} =- \beta^e I
\end{equation}{}
\begin{equation}
\frac{dI}{dt} =\beta^e I - \alpha I
\end{equation}{}
\begin{equation}
\frac{dT}{dt} = \alpha I -\gamma T  
\end{equation}{}
\begin{equation}\label{eq:SITR2}
\frac{dR}{dt} = \gamma T
\end{equation}{}

The parameter $\beta^{e}_t = \beta\frac{S_{t}}{N_{t}} $ is the real transmission rate over time, taking into account the total population size $N$ as in the SIR model. Assuming all the parameters $\boldsymbol{\theta} = [\beta^{e}, \alpha, \gamma]$ time dependent, the SITR model in equations \eqref{eq:SITR1}-\eqref{eq:SITR2} can be discretized with respect to the time variable, giving the following equations:
\begin{equation}\label{eq:St}
S_{t+1} = S_{t} - \beta_{t}^e I_{t}
\end{equation}{}
\begin{equation}
I_{t+1} = I_{t}+\beta_{t}^e I_{t} - \alpha_{t}I_{t}
\end{equation}{}
\begin{equation}
 T_{t+1} = T_{t}+\alpha_{t}I_{t}-\gamma_{t}T_{t}  
\end{equation}{}
\begin{equation}\label{eq:Rt}
R_{t+1} = R_{t} + \gamma_{t} T_{t} 
\end{equation}{}
%Where $S$ denotes the susceptible population size, $I$ the infected people who are not isolated from the population and $R$ the recovered population. 
which is a linearized approximation of the original SIR model with the additional compartment $T$. This provides the model prediction of the compartment states $\mathbf{X}=[S, I, T, R]^T$ given all parameters including $\beta^e$. The other variables are the same as in the SIR model, where $S$ denotes the susceptible population, $I$ the infected people who are not isolated from the population. 
%$T$ denotes the number of treated people, which is observable since this is the the number of confirmed cases minus the number of recovered cases. 
The parameters $\gamma$ and $\alpha$ denote the recovery and transition rate given by total of incubation and admission days.  %The state vector is then given by $\textbf{x}^{T} = [S,I,T,R]$. 
To extend the model and incorporate information not just of the last timestep, we introduce a model extension which bases model predictions on a sliding window of length $\tau$, similar to a 4D-VAR approach \cite{asch_2016}.
For a given time window $[t+1, t+\tau]$ and assuming to have observations of the variable $T_t$ which we denote here with $T^{obs}_t$, the resulting assimilation scheme is given by
%Original equation 10082020
% \begin{equation}
%     %I^{DA}_{t}  = \operatorname*{argmin}_{I} =\sum^{t+\tau}_{i=t+1} ||\textbf{H}(i)-\hat{\textbf{H}}(i|I,\beta^{e}_{t-1})||_{\textbf{Q}^{-1}_t} + ||I-\hat{I}_{t}||_{\textbf{P}^{-1}_{t}}
%      I^{DA}_{t}  = \operatorname*{argmin}_{I} \sum^{t+\tau}_{i=t+1} ||T^{obs}_t-H(i|I,\beta^{e}_{t-1})||_{\textbf{Q}^{-1}_t} + ||I-I_{t}||_{\textbf{P}^{-1}_{t}}
% \end{equation}{}

\begin{equation}\label{eq:1}
J(I) =  \sum^{t+\tau}_{i=t+1}||T^{obs}_{i}-T^{pred}_{i}||_{\textbf{Q}^{-1}_{t}} + ||I-I^{pred}_{t}||_{\textbf{P}^{-1}_{t}}%H(i|I,\beta^{e}_{t-1})
\end{equation}
and 
\begin{equation}\label{eq:3a}
\mathbf{X}^{pred}_{t+1}=F(\mathbf{X}^{}_t, \boldsymbol{\theta}_{t})
\end{equation}

\begin{equation}
    %I^{DA}_{t}  = \operatorname*{argmin}_{I} =\sum^{t+\tau}_{i=t+1} ||\textbf{H}(i)-\hat{\textbf{H}}(i|I,\beta^{e}_{t-1})||_{\textbf{Q}^{-1}_t} + ||I-\hat{I}_{t}||_{\textbf{P}^{-1}_{t}}
     I^{DA}_{t}  = \operatorname*{argmin}_{I}  J(I)
\end{equation}{}
which in a first step infers the number of infected people $I$. To estimate the infection rate, in a second step we minimize
\begin{equation}
  %  \beta_{t}^{e}  = \operatorname*{argmin}_{\beta^{e}} =\sum^{t+\tau}_{i=t+1} ||\textbf{H}(i)-\hat{\textbf{H}}(i|I,\beta^{e}_{t-1})||_{\textbf{Q}^{-1}_t} %+L(\beta^{e})
     \beta_{t}^{e}  = \operatorname*{argmin}_{\beta^{e}} \sum^{t+\tau}_{i=t+1} ||T^{obs}_i-T^{pred}_i||_{\textbf{Q}^{-1}_t}
\end{equation}{}
% \begin{equation}
%     \beta_{t}^{e}  = \operatorname*{argmin}_{\beta^{e}} =\sum^{t+\tau}_{i=t+1} ||\textbf{H}(i)-\hat{\textbf{H}}(i|I,\beta^{e}_{t-1})||_{\textbf{Q}^{-1}_t}+L(\beta^{e})
% \end{equation}{}

%original equation 10082020
% which in a first step infers the number of infected people $I$ and where $H:I\to T$ is a linear transformation function usually called observation function \cite{asch_2016}. To estimate the infection rate, in a second step we use minimize
% \begin{equation}
%   %  \beta_{t}^{e}  = \operatorname*{argmin}_{\beta^{e}} =\sum^{t+\tau}_{i=t+1} ||\textbf{H}(i)-\hat{\textbf{H}}(i|I,\beta^{e}_{t-1})||_{\textbf{Q}^{-1}_t} %+L(\beta^{e})
%      \beta_{t}^{e}  = \operatorname*{argmin}_{\beta^{e}} \sum^{t+\tau}_{i=t+1} ||T^{obs}_t-H(i|I,\beta^{e})||_{\textbf{Q}^{-1}_t}
% \end{equation}{}

which updates $\beta$ conditioned on assimilated values of $I$.
The resulting algorithm implements a 4D-VAR assimilation scheme in cost function \eqref{eq:1}, where forecasts and parameter estimates are based on a sliding window over time.
Without preconditioning, the algorithm updates the model parameter values with the noise and observation matrices $\textbf{Q}$ and $\textbf{P}$ being fixed hyperparameters. 
In order to present results which may have major policy implications, correct and robust estimation of initial conditions and hyperparameters is of high importance, we therefore introduce a formalization and preconditioning of the covariance matrices $\textbf{Q}$ and $\textbf{P}$ before applying the assimilation scheme, which is named hybrid data assimilation.
%In order to introduce a more rigorous determination of the covariance matrices, we next discuss the introduction of a hybrid assimilation scheme which makes the SITR model more robust to initial choices of the matrices. 

% Equation~\eqref{eq:3} can be linearised about the background state vector which will result in the incremental form of the cost function given by
% \begin{equation}\label{eq:4}
%     J(\delta\textbf{x})=\frac{1}{2}(\delta\textbf{x})^{T}\textbf{B}^{-1}(\delta\textbf{x})+\frac{1}{2}\sum_{k=0}^{K}(\delta\textbf{d}_{k})^{T}\textbf{P}_{k}^{-1}(\delta\textbf{d}_{k})
% \end{equation}
% where where $\delta\textbf{d}_{k}$ = $\textbf{d}_{k}-\textbf{H}_{k}\delta\textbf{x}$ and $\textbf{d}_{k} = \textbf{y}_{k}^{o}-\textbf{H}_{k}\textbf{x}_{k}^{b}$. This cost function is minimised to provide a solution,
% \begin{equation}
%     \delta\textbf{x}^{a} = argmin\thinspace J(\delta\textbf{x})
% \end{equation}
% and 
% \begin{equation}
%     \textbf{x}^{a} = \textbf{x}^{b}+\delta\textbf{x}^{a}
% \end{equation}
% with $\textbf{x}^{a}$ denoting the state vector after data assimilation.

% This form is called 3D First Guess at Appropriate Time or 3D-FGAT \cite{lorenc_2000} and is useful applications where the model operator, $\boldsymbol{\mathcal{M}}$ is not available.

% \subsection{Sensitivity Analysis}
% Add table with results

\section{Uncertainty in Infection Rates}
\begin{figure}[!htpb]
\centering
 \includegraphics[width=0.49\linewidth]{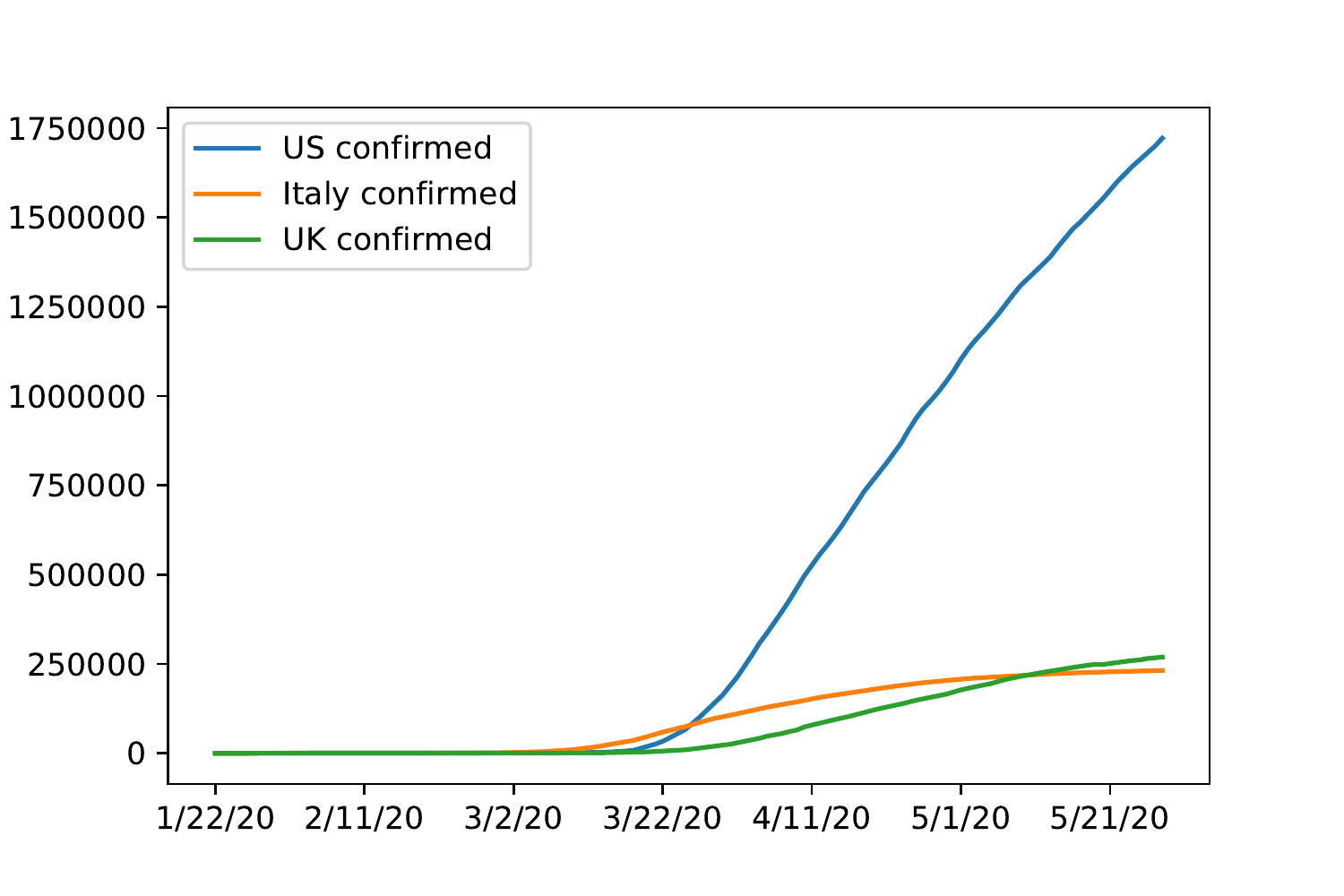}
  \includegraphics[width=0.49\linewidth]{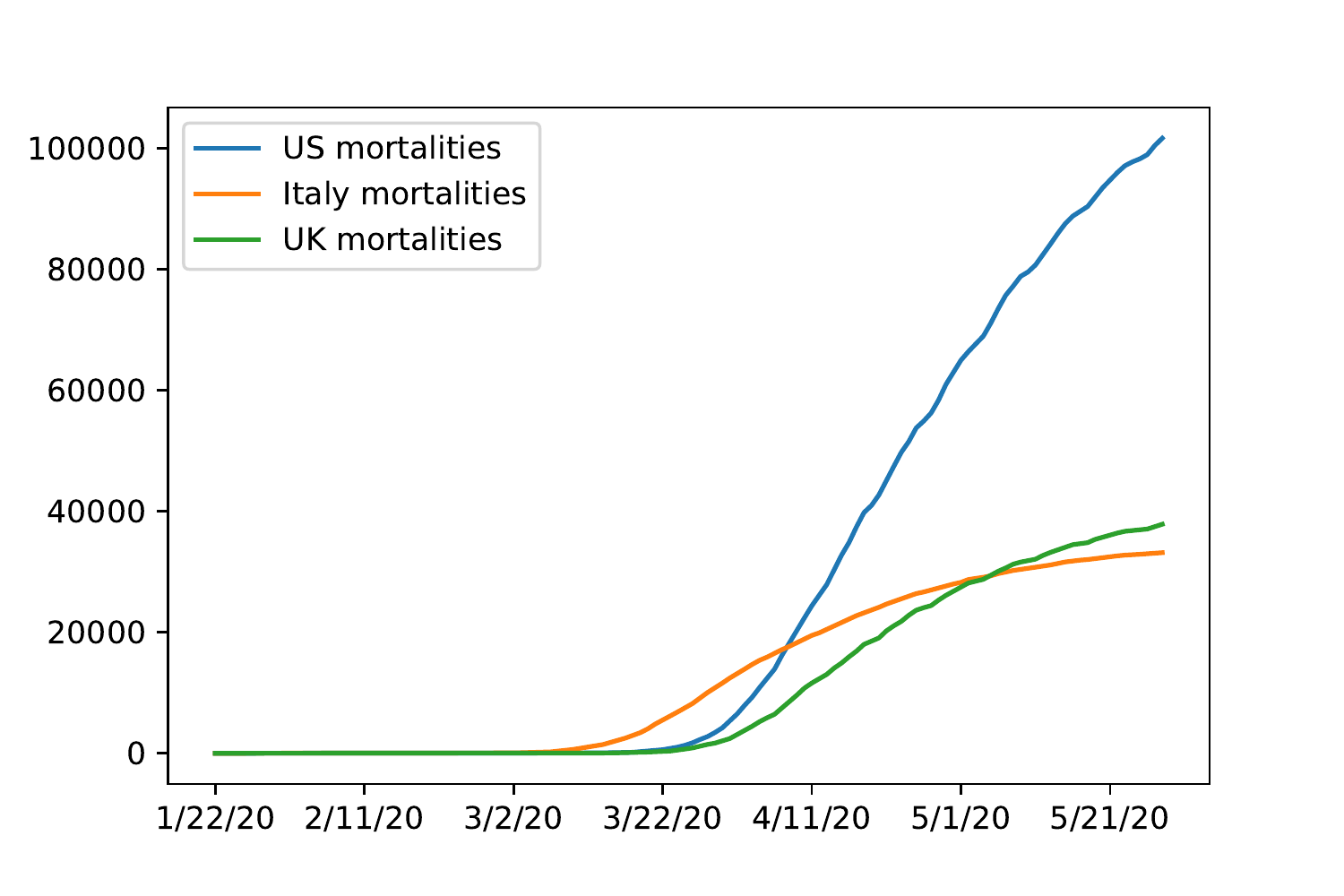}
  \caption{Number of mortalities and confirmed cases for the United States, Italy and the United Kingdom}\label{fig:bj}
\label{fig:global_data1}
\end{figure}

The data that is being observed is highly aggregated and suffers from uncertainty firstly due to human measurement error and secondly due to number of confirmed cases being a noisy subset of the true number of infections. Furthermore different definitions for confirmed cases or the cause of mortalities due to Covid19 in various countries adds noise and uncertainty to the data. The figures in Fig.~\ref{fig:global_data1} show example data of confirmed cases and mortalities.
This uncertainty in the data mandates a methodology that incorporates the uncertainty into the model. The DA-SIR model takes this uncertainty into account via the values of the covariance matrices $\textbf{Q}$ and  $\textbf{P}$. \newline
The state and observation covariance matrices  $\textbf{Q}$ and  $\textbf{P}$ determine the weight of new observations when updating the parameters of the model. As is stipulated in the cost function in Eq.~\ref{eq:3}, the inverse of the covariance matrices are used to weight the terms of the observation and model operator. Thus very noisy data yielding large error covariance matrices will result in less weight of the term containing the model operator, i.e. less certainty is put on the data.\newline
Therefore the next section studies different covariance matrix setups to take data uncertainty into account. We include detailed steps for the computation of the covariance matrix in the appendix, where we outline the ensemble variational approach applied to generate robust covariance matrix estimates.

%PN: Human measurement error in the data.
%Q and P are weights for loss function, this takes into account uncertainty of data.
%We follow up with a sensitiviy analysis, and appendix exact procedure for Q P.
\subsection{Sensitivity Analysis}
As we mentioned in Section 4.1, the choice of the covariance matrices strongly affect the efficiency and the accuracy of the assimilation approach. As the available data is not accurate enough, in order to justify our estimations, we run a sensitivity analysis to study the impact of our estimated parameters and covariance matrices into the model predictions, using a subset of the data as illustration.
To illustrate the hyperparameter sensitivity we compare the number of estimated infected people and we apply a mean root squared forecasting error (MRSFE) metric:
\begin{equation}
     MRSFE= \sum_{n=0}^{N}\bigg( \frac{\sum_{\tau=\tau_0}^{T-h}\sqrt{(y_{t,n}^r-\hat{y}_{t,n})^2}}{T-h-\tau_0+1} \bigg)
 \label{eq:MSFE}
\end{equation}{}
where $\hat{y}_{t,n}$ represents the model prediction, $y_{t,n}^r$  the real observation with forecast horizons defined by $h=1$, and $\tau_0=1$ the starting period of the forecast for $n$ variables. 
The results are given in table \ref{tab:table_sensitivity}.

The results in the table show that a naive setup using unit covariance matrices leads to detrimental fit of the model with relatively high forecasting errors compared to other covariance combinations, with generally better performance for high values for observation covariance matrix $\textbf{P}$, meaning that prediction accuracy increases when less weight is put on the model dynamics and more weight on the observations. The lowest forecasting error for $T$ is given with a unit observation matrix  $\textbf{P}$ and an error covariance matrix  $\textbf{Q}$ of 10, showing that the optimal combination of covariance matrix values is non-linear and requires a carefully considered estimation algorithm, as our proposed ensemble approach. \newline 
The computation of the covariance matrices is performed via a hybrid-assimilation approach where the covariance matrices are estimated using an ensemble approach. In this approach, the values of the covariance matrices are generated through sampling multiple synthetic trajectories using the SITR which is initialized with different parameters for each draw as well as calculating the empirical residual covariance matrix for the data. The details of the procedure are outlined in the appendix.
%\begin{center}
\begin{table}[h]
\centering
% \label{features}
\begin{tabular}{l|c|c|c|c|c|c}
%\toprule
\hline
%\midrule
%Treatment &      230760&        2.31336077e+05 &     230792          &      231517 &      230782 &    230934   \\
%Infections &     2.10311881e+05 &        2.10212634e+05 &    2.10254953e+05  &      2.10243172e+05 &     2.10257659e+05  &    2.10260399e+05   \\
$\mathbf{P}$ Value   &   0.1         &         1 &         1                &    1         &    100 &    100     \\
$\mathbf{Q}$ Value   &     0.5       &      100 &           1              &     10        &      1       &         10    \\
\hline
Treatment &      230760&        231336 &     230792          &      231517 &      230782 &    230934   \\
\hline
Infections &     210311 &        2102126 &    2102549  &      2102431 &     2102576  &    2102603   \\
\hline
MRSFE T &     888&                      863            &           901    &           807 &      829 &      836   \\
\hline
MRSFE R &      39476              &         39291 &           39431 &                 39381 &      39459 &      39469   \\
\hline
%\bottomrule
\end{tabular}
\caption{Sensitivity analysis for different values of observation and model error covariance matrices. The first two rows show number of latent infected patients and patients under treatment predicted for the 28.05.2020, the last day in the sample. The last two rows show the mean forecasting errors for treated patients and confirmed cases over the full sample for each covariance matrix configuration. The table exemplifies a bad fit with a high amount of forecasting errors when using a naive unit covariance setup.}
\label{tab:table_sensitivity}
\end{table}{}
%\end{center}{}

\begin{figure}[!htpb]
\centering
 \includegraphics[width=0.49\linewidth]{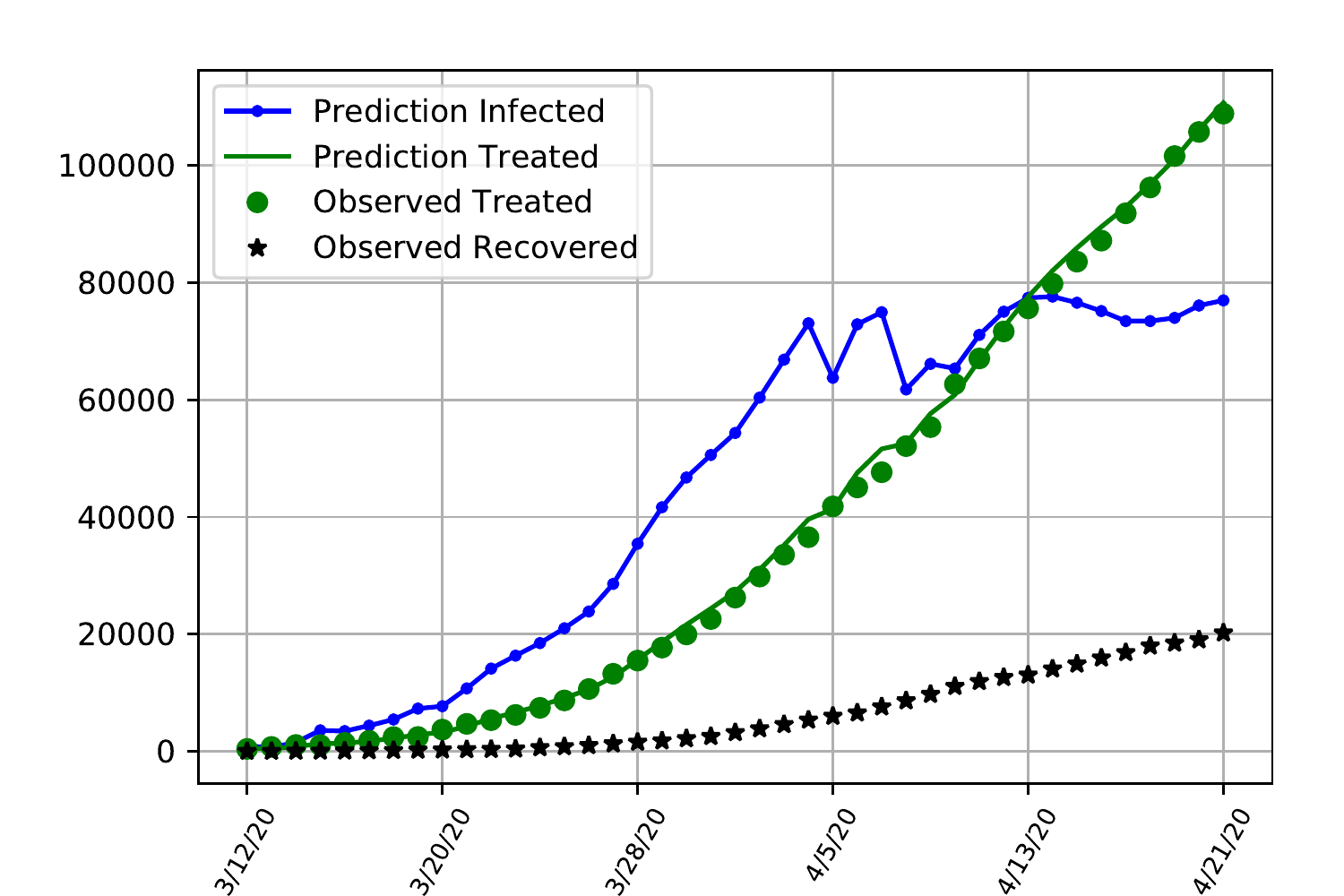}
  \includegraphics[width=0.49\linewidth]{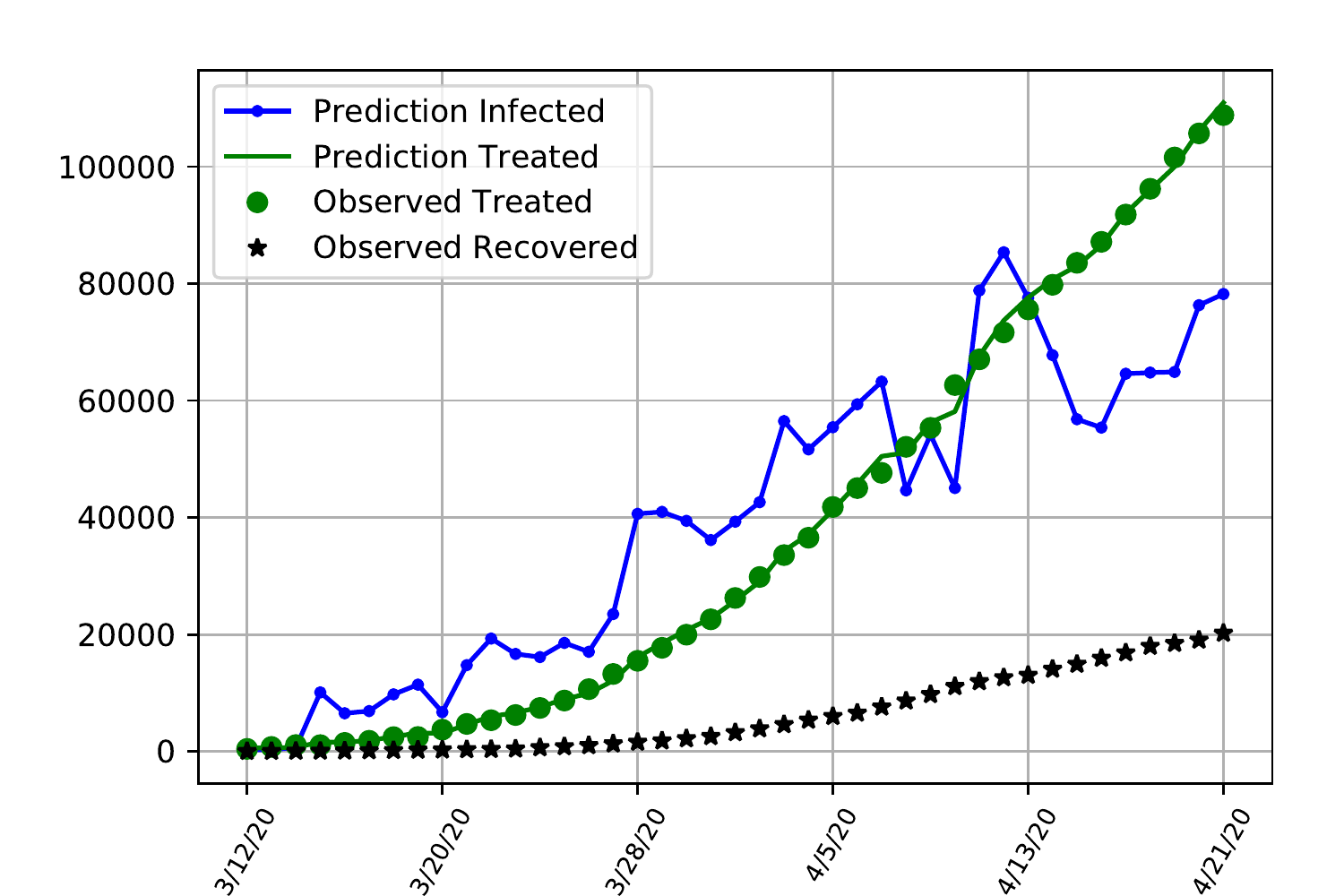}
  \caption{Number of infected and treated cases for the United Kingdom, depicting the difference between infection numbers for robust (left) and naive unit covariance matrices (right)}\label{fig:bj}
\label{fig:sensitivity1}
\end{figure}

Fig.~\ref{fig:sensitivity1} depicts different infection curves given the naive unit covariance setup on an example time period of the data in table \ref{tab:table_sensitivity} and show that the dynamics are affected by the choice of the covariance matrices. 
The updated model assimilates new observations of infected patients and people recovered from the virus. The long run dynamics predict a recent spike in the number of infected people in the United Kingdom. The total number of people being treated in hospitals follows with a small lag and is still growing towards the end of the example set.\newline
The trajectories for patients under treatment are similar but the dynamics for latent infections contain some discrepancies. The left hand side depicts the trajectory of estimated latent infections and patients over time using the ensemble approach with robust covariance matrices, whereas the right hand figure depicts results using a simple unit covariance setting. The left hand side depicts a much smoother and more well-behaved series of treated cases as well as infections, approximating a more reasonable stable growth path, whereas the unit covariance case would yield much more erratic, unrealistic fluctuations in infection numbers over time, with many sudden drops in infection numbers.\newline
Since the number of treated patients is observable, we use generated forecasts of treated persons as a forecasting metric to evaluate the model fit.
The tables \ref{tab:table_unitcov} and \ref{tab:table_hybridcov} give excerpts from the forecasts values of infected and treated patients as well as the corresponding MRSFE for treated patients in hospitals.
For the unit covariance table \ref{tab:table_unitcov} it is observable how the unrealistic dip in forecasted infections which is visible in the right side plot in Fig.~\ref{fig:sensitivity1} causes a large spike in forecasting errors for treated people beginning of April onwards.
Comparing it with a hybrid assimilation approach in table \ref{tab:table_hybridcov} reveals an overall lower number of forecasting error and better fit.
The sensitivity of results confirm the need for a more rigorous algorithm of covariance estimates given the few and noisy datapoints.

\begin{figure}[!htpb]
\centering
 \includegraphics[width=0.49\linewidth]{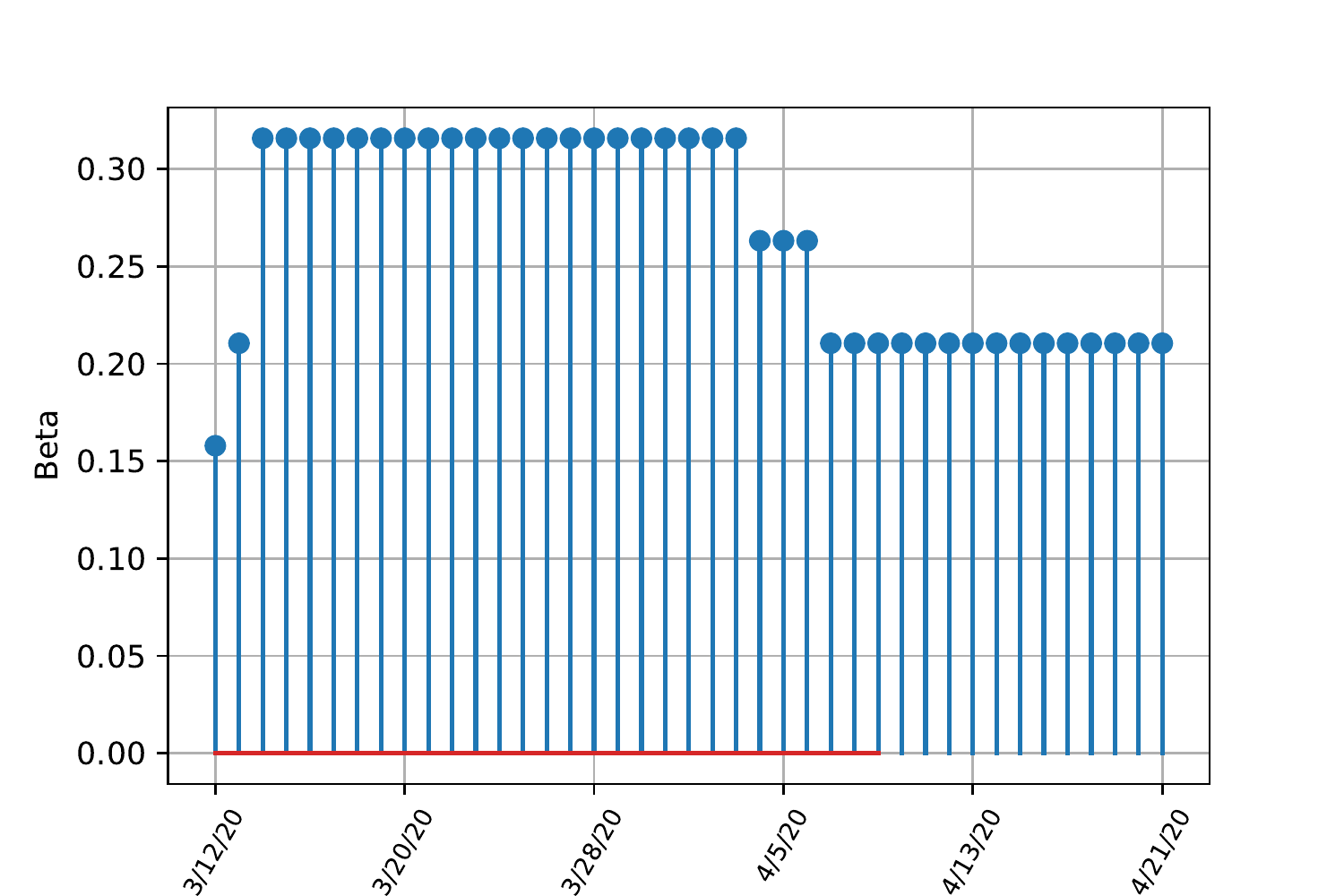}
  \includegraphics[width=0.49\linewidth]{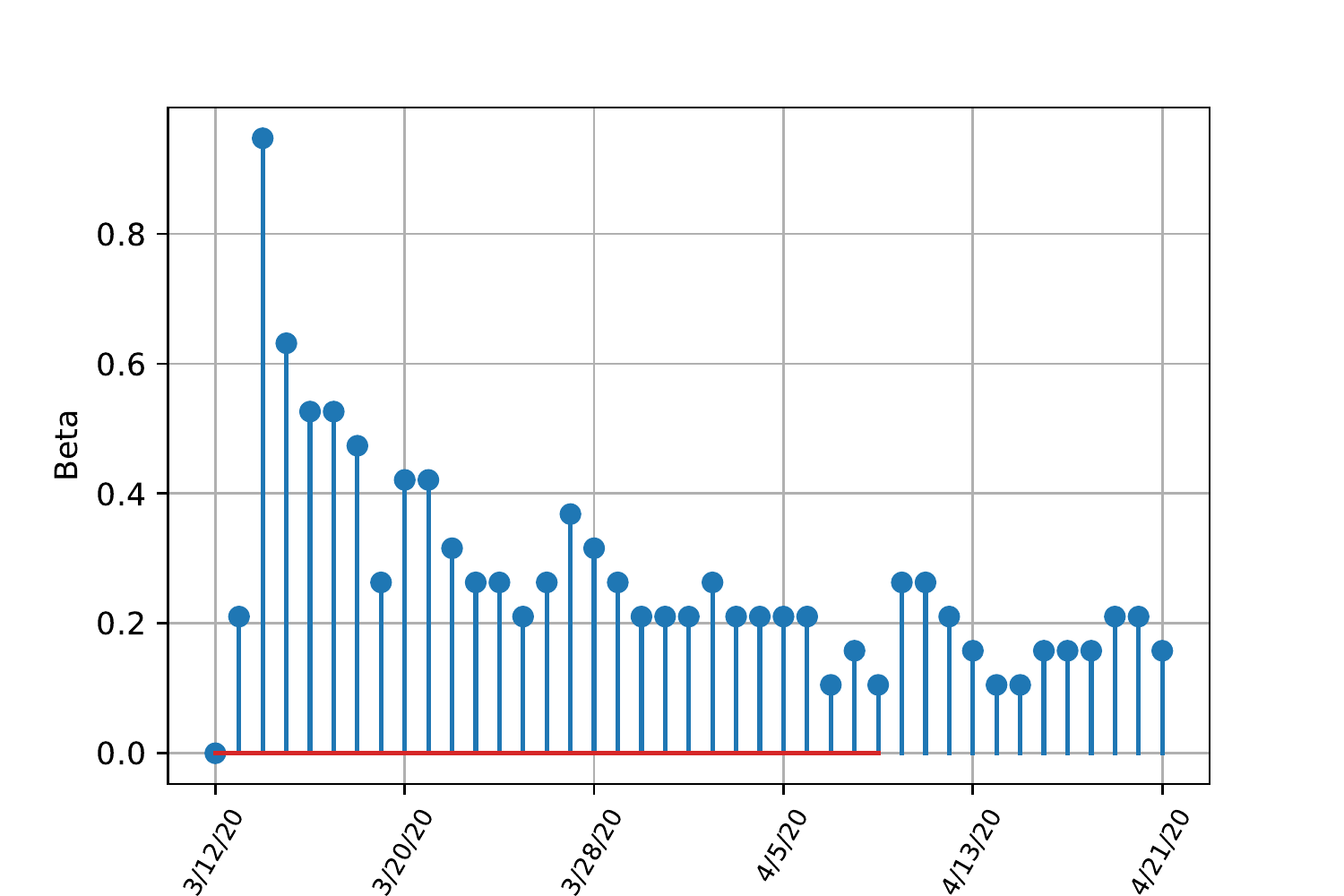}
  \caption{ Estimates for transmissibility rates beta for the United Kingdom, depicting the difference between infection numbers for robust (left) and unit covariance matrices (right)}\label{fig:bj}
\label{fig:cov_comparison}
\end{figure}

Comparing the left and right bottom figures of Fig.~\ref{fig:cov_comparison}, the transmissibility rate $\beta$ shows less variation over time, which differs from the model without ensembles where strong variation is visible. Both estimates depict the downward trend of transmissibility. The high variability of the unit covariance matrix estimate implies that the transmissibility is affected more easily by external factors which change the dynamics of new infections. Thus the robust model estimates imply that, within the sample period, the transmission rate is more stable and unaffected by changes in observations because a uncertainty weight is given by the covariance matrix estimates. %This is also showing how the infection rate is decreasing over time towards the end of the example. 
We next proceed to apply out methodology to the full dataset analysing the US, the UK and Italy.

%\begin{center}
\begin{table}[]
\centering
% \label{features}
\begin{tabular}{l|c|c|c|c|c|c}
%\toprule
\hline
Date & 03-09 & 03-16 & 04-12 & 04-19 & 05-21 & 05-28  \\
\hline
%\midrule
Infected Patients    &     21 &    85 &     12629 &    18492 &    36042 &    37837     \\
Treated Patients     &      300 &     1458 &     71650 &     101575 &    214866 &    231290     \\
RSFE Treated &       261 &  110 & 2049 & 1592 & 1308 &  498 \\
RSFE Infections & 43 & 114 & 4439 & 13950 & 133162 & 172417 \\
%RSFE Infections & 4.30000e+01 & 1.14000e+02 & 4.43900e+03 & 1.39500e+04 & 1.33162e+05 & 1.72417e+05 \\
\hline
%\bottomrule
\end{tabular}
\caption{Selected data points for predicted number of infected and treated patients, as well as the MRSFE. Results are obtained using a naive unit covariance matrix.}
\label{tab:table_unitcov}
\end{table}{}
% For unit
% T_MRSFE : 901.5301204819277
% R_MRSFE : 39431.9156626506

\begin{table}[]
\centering
% \label{features}
\begin{tabular}{l|c|c|c|c|c|c}
%\toprule
\hline
Date & 03-09 & 03-16 & 04-12 & 04-19 & 05-21 & 05-28  \\
\hline
%\midrule
%3.13534632e+01, 1.54320308e+02, 1.67555215e+04, 3.20230780e+04,1.68831181e+05, 2.09999354e+05
Infected Patients    &     31 &    154 &     16755 &    32023 &    168831 &    209999   \\
Treated Patients     &      290 &     1405 &    72363 &     101051 &    214629 &    232074     \\
RSFE Treated &       10&  53 & 713 & 524 & 237 &  784 \\
RSFE Infections & 10 & 69& 4126 & 1353 & 132789& 172162 \\
%RSFE Infections & 4.30000e+01 & 1.14000e+02 & 4.43900e+03 & 1.39500e+04 & 1.33162e+05 & 1.72417e+05 \\
\hline
%\bottomrule
\end{tabular}
\caption{Selected data points for predicted number of infected and treated patients, as well as the MRSFE. Results are obtained using the hybrid assimilation covariance matrix.}
\label{tab:table_hybridcov}
\end{table}{}
%for hybrid
% T_MRSFE : 939.9759036144578
% R_MRSFE : 39167.5421686747

% For unit
% T_MRSFE : 901.5301204819277
% R_MRSFE : 39431.9156626506

\section{Empirical Results}

\subsection{Trend Analysis of International Data}
\begin{figure}[!htpb]
\centering
 \includegraphics[width=0.6\linewidth]{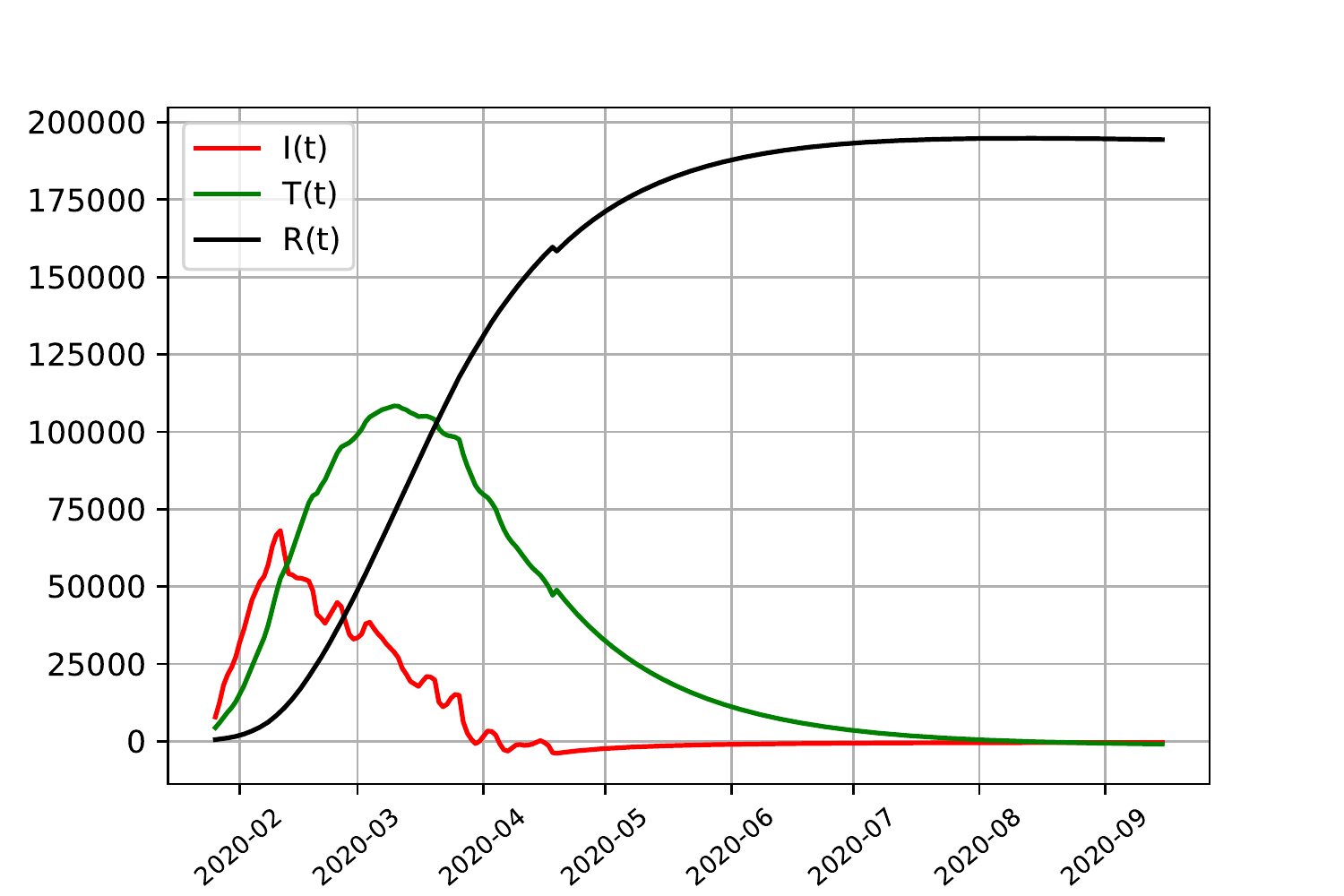}
  \caption{SITR results for Italy, showing estimates of latent infections (red), recoveries (black) and hospitalizations (green)}
\label{fig:SITR_it_long}
\end{figure}

To illustrate the flexibility of our approach we apply our analysis to an international comparison with additional results for the United States, the United Kingdom as well as Italy. 
%Since country level data is more readily available than city specific data but is less reliable than more granular city-level data such as patients under treatment. We conduct experiments using both the DA-SIR model as well as the DA-SITR model. 
\newline For the models we focus on the number of infected people extrapolated from the number of confirmed cases and recoveries. The data was obtained from the John Hopkins University Coronavirus Resource Center\footnote{https://coronavirus.jhu.edu/map.html}. We compare data and show test results for different forecast horizons and parameter estimates. Given the confirmed coronavirus cases we infer the amount of infected people and do forecasts to estimate the approximate development of the epidemic. We estimate the model on a sample of daily data until the 28/05/20 and evaluate models based on their fit and infection curves.\newline
Although long-term forecasts are of limited use in fast changing scenarios such as the current pandemic, they nevertheless can provide rough guidance on a potential peak of infection rates.
Comparing cases for all three countries Fig.~\ref{fig:SITR_it_long} depicts the long-run dynamics of the epidemic in Italy, where according to the model the peak of infections has already occurred in march, with a gradual decrease in infection numbers afterwards. The absolute number of patients under treatment decrease throughout April and May.

%Forecasting errors on a nationwide level are higher compared to city based estimates in the previous section due to higher absolute values, with 
%the predicted confirmed cases deviating most visibly from the observations during the peak of infections.\newline
%According to the SIR specifications a high absolute amount of infections affects the marginal increase in new infections, thus the prediction of confirmed cases increases as can be seen from the shape of the black solid line in the figure. For nationwide estimates, constrained testing capabilities to confirm cases are unlikely to increase with the rate of infections, thus the difference in predicted and confirmed cases can be due to limited testing capabilities.
%In the data assimilation correction cycle forecasting errors are feed back into the model and lead to a correction of parameters which leads to model adjustments and decreases in predicted infections and confirmed cases over time, eventually reducing the gap in forecasted and observed values.\newline
%According to the model estimates, the infection rate at the end of the sample is around 10 percent of the population, which is forecasted to drop to 6 percent in a 5 day ahead forecast. To put the depicted infection numbers in both countries into perspective, the absolute values of the forecasted infection numbers are also given in table \ref{tab:table_usit}.\newline

\begin{figure}[!htpb]
\centering
 \includegraphics[width=0.48\linewidth]{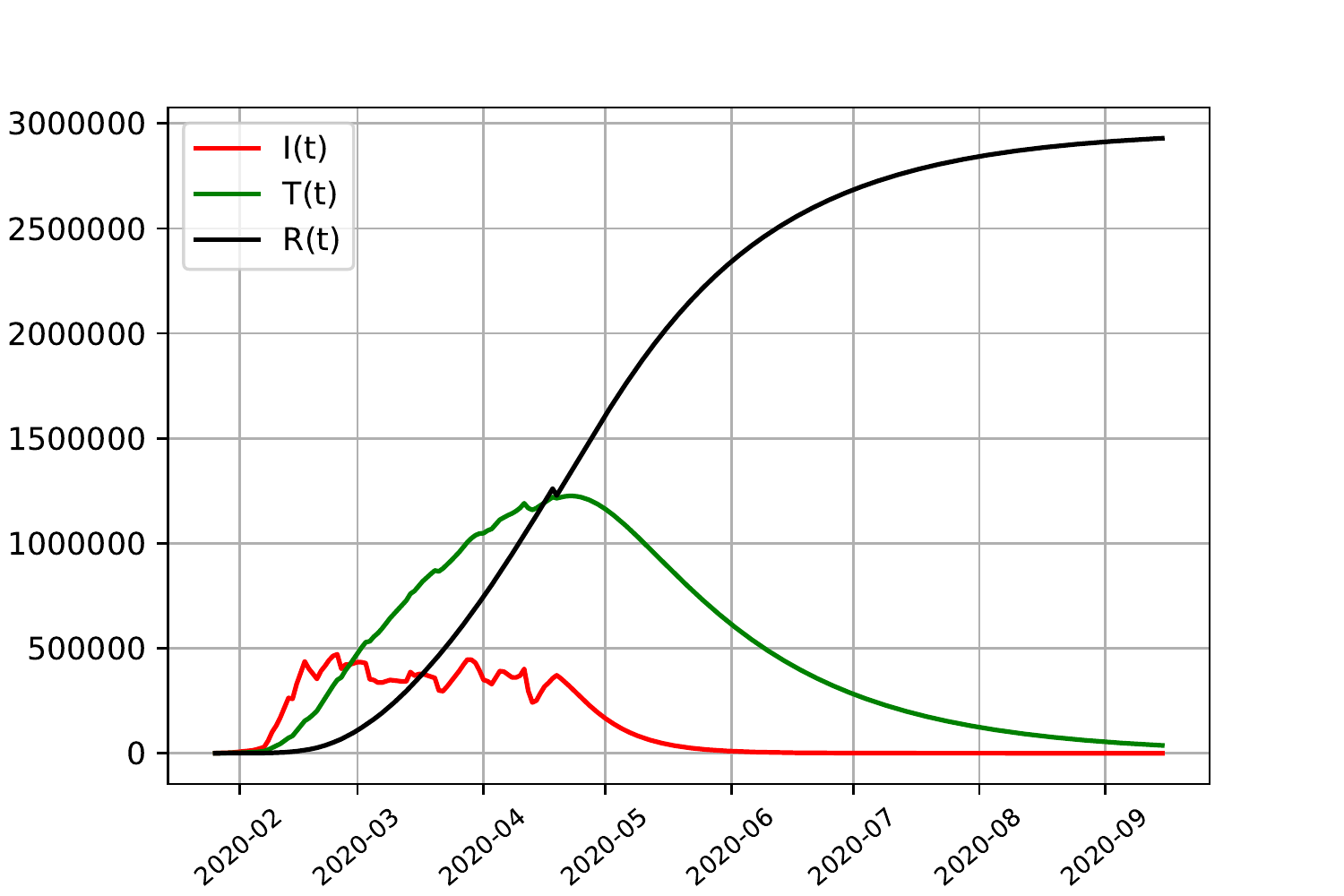}
  \includegraphics[width=0.48\linewidth]{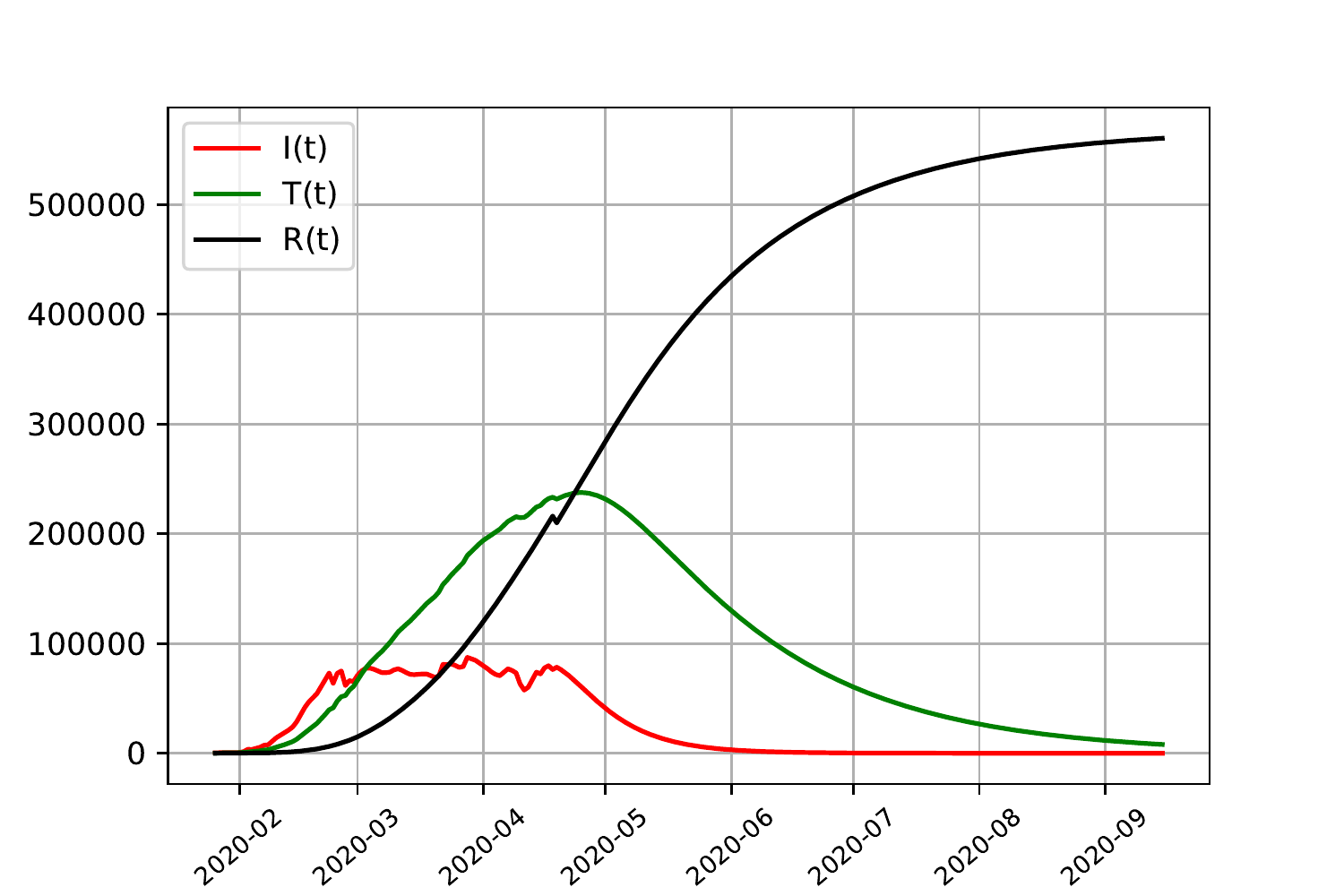}
  \caption{SITR results for the US (left) and UK (right), showing estimates of latent infections (red), recoveries (black) and hospitalizations (green)}\label{fig:bj}
\label{fig:SITR_US}
\end{figure}

Comparing Italian results with the United States and the United Kingdom in Fig.~\ref{fig:SITR_US} shows that the trajectories of Italy differs from both countries, with the UK and the US showing similar patterns. When infection numbers in Italy already peaked, the number of latent infections depicted in red has not decreased, but has merely stabilized at a high level, whereas the hospitalized cases under treatment are still growing, with a forecasted peak by end of April, which contrasts to the Italian peak in March. The number of infections are increasing within sample, as is the forecasted number of infections. The peak of latent infections stabilses at slightly below half a million cases in the US and 100,000 cases in the UK. For both countries the total number of infections tappers off by the end of September, whereas this already happens by July for the case of Italy.

The different levels of infections are likely due to different inception dates of the pandemic, having started earlier in Italy than the United States, with an eventual peak of the United States not visible yet given the current data sample.%, as well as a high amount of uncertainty given very limited testing capabilities in Italy and especially the United States.\newline
The results indicate that the pandemic has reached a peak in Italy recently, the dynamics for the United Kingdom and especially the United States indicate that no plateau has been reached yet and that the number of infections is likely to increase.\newline

%The predicted confirmed cases are closer to observed values due to percent wise relatively low infection numbers compared to Italy, although the model predicts an increasing trend given current model dynamics. Since a lower amount of the total population is infected, forecasts of confirmed cases are affected less by the high number of infected cases as the SIR dynamics imply and is visible for Italy.\newline

\subsection{Short Term Dynamics}

\begin{figure}[!htb]
\centering
 \includegraphics[width=0.5\linewidth]{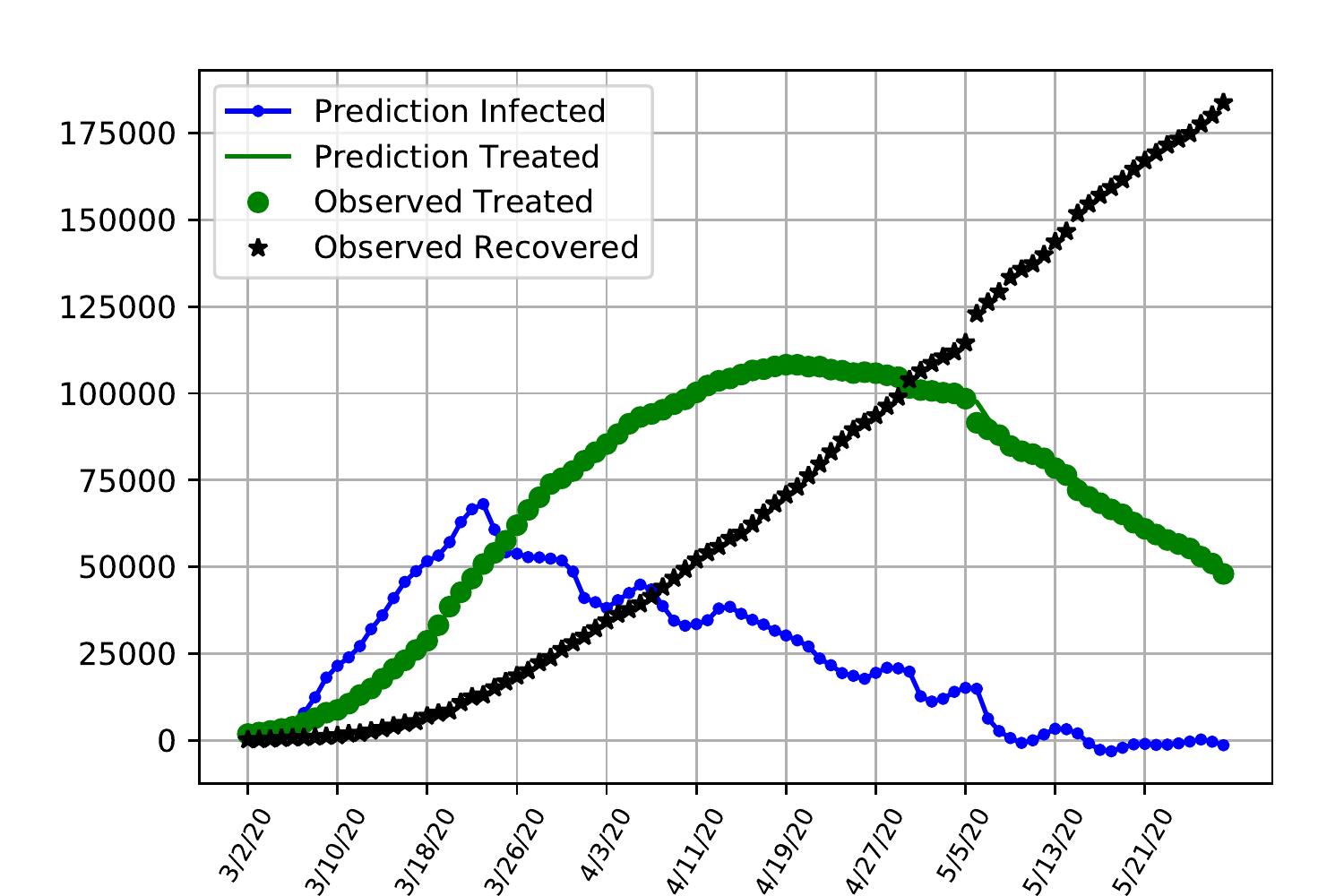}
  \caption{SITR short run dynamics Italy, showing recoveries (black), latent infections (blue) as well as observed and predicted numbers of treated patients (green)  }\label{fig:it_short}
\label{fig:SITR_it}
\end{figure}

\begin{figure}[!htb]
\centering
 \includegraphics[width=0.5\linewidth]{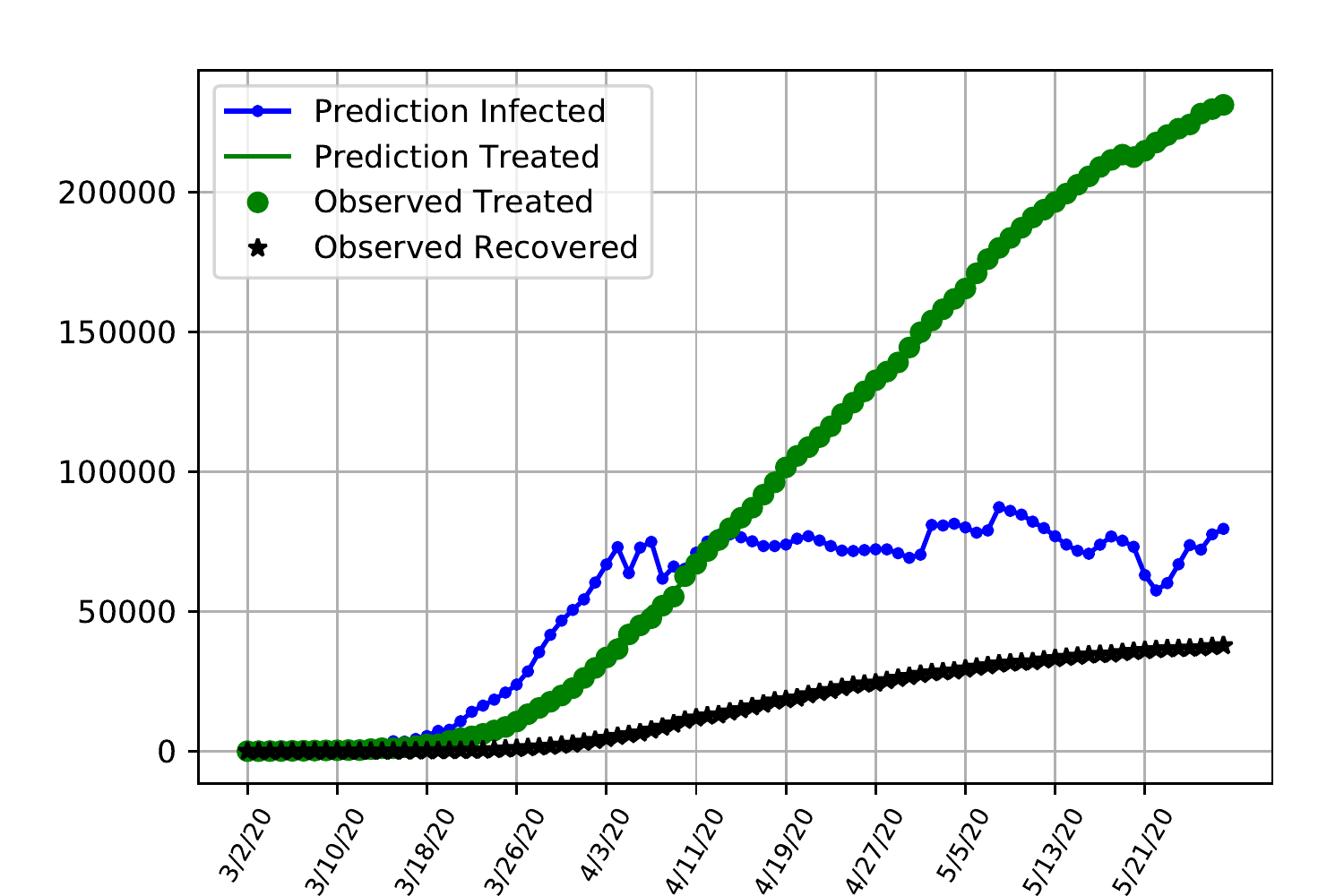}
  \caption{SITR short run dynamics in the United Kingom, showing recoveries (black), latent infections (blue) as well as observed and predicted numbers of treated patients (green)}\label{fig:uk_short}
\label{fig:SITR_uk}
\end{figure}

\begin{figure}[!htb]
\centering
 \includegraphics[width=0.5\linewidth]{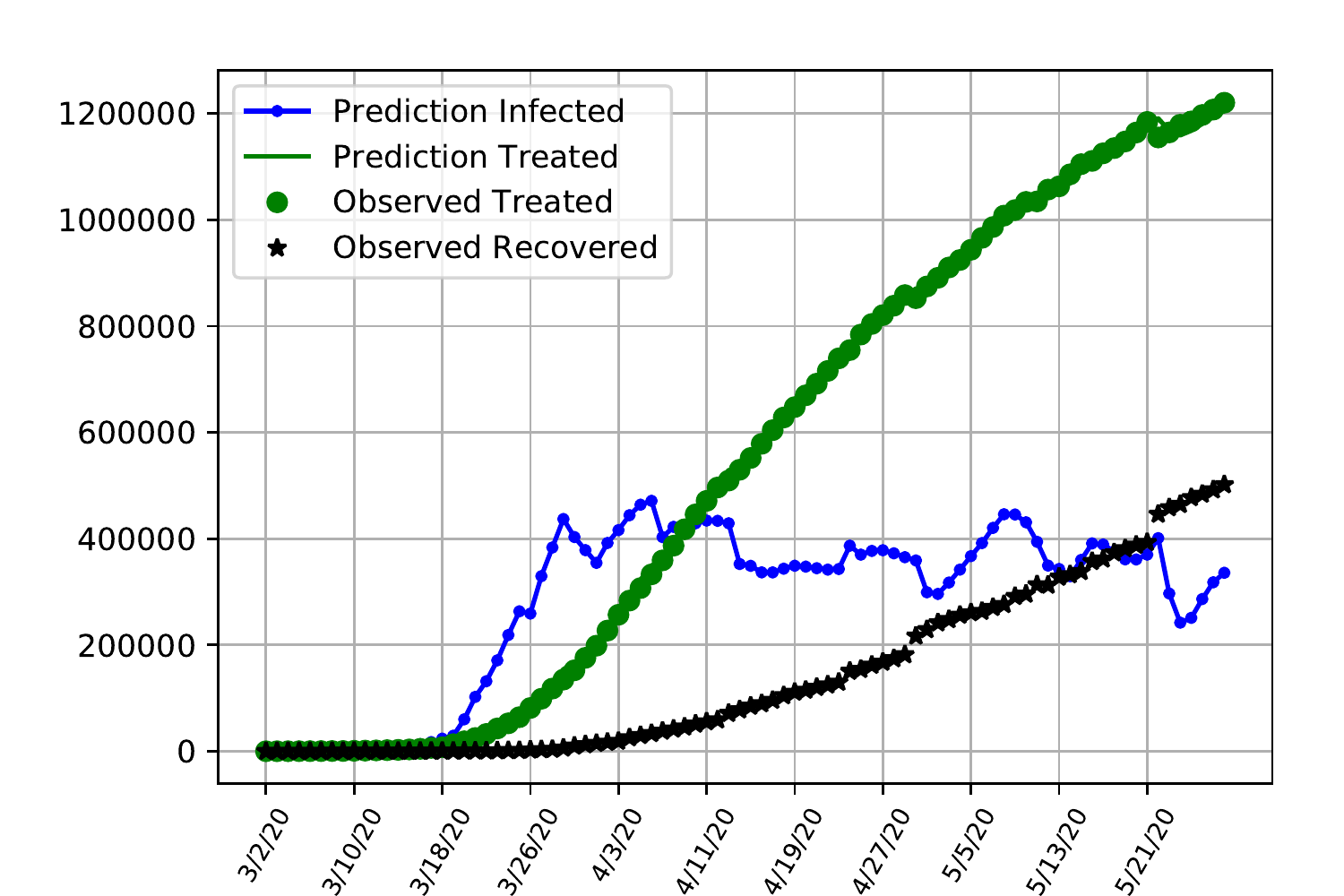}
  \caption{SITR short run dynamics in the United States, showing recoveries (black), latent infections (blue) as well as observed and predicted numbers of treated patients (green) }\label{fig:us_short}
\label{fig:SITR_us}
\end{figure}

The results given by the dynamic SITR model highlight the different phases of development in Italy and the United Kingdom and the United States, as is depicted in Fig.~\ref{fig:SITR_it}, Fig.~\ref{fig:SITR_uk} and Fig.~\ref{fig:SITR_us} respectively. The figures depict the predicted latent infection numbers, observed and predicted hospital treatment numbers as well as the number of recoveries. We also provide accompanying tables where we report the MRSFE fit of observed confirmed and treated cases. We first discuss the trajectories of infections and follow up with a discussion on the parameter estimates in the next section.
%Results align with the previous analysis, with the absolute numbers being lower in magnitude because they are distributed across more compartments.\newline 

Results align with the previous analysis where in Fig.~\ref{fig:SITR_it} it is observable that in Italy the number of latent asymptomatic cases has decreased with the majority of patients being hospitalized with a decreasing trend for both latent infections and hospitalized patients. The pattern of latent infections follows a shaped curve, with the majority of infections already peaked in march and now being on a trajectory exhibiting signs that the pandemic is under control. 
Table \ref{tab:table_hybridcov_it} provides additional details for the model performance, showing the number of latent infections and treated patients in the model. We show the forecasting errors when fitting the model in order to be able to compare model fit between all three countries. The forecasting errors depict how the model fits the data, with the estimates of treated patients performing particularly well compared to the UK and US (tables \ref{tab:table_hybridcov_uk} and \ref{tab:table_hybridcov_us}). At the end of the sample on the 28th of May the number of latent infected patients is slightly above 180,000 and the number of patients under treatment is below 48,000, with the values exemplifying the decrease from the previous month.\newline
Overall number of infected but not hospitalized patients increases initially and are starting to trend downward after a peak at the end of March.\newline

The hospitalization numbers are displaying a curve like behaviour, following the infections with a lag. The maximum number of patients under treatment are reached in the middle of April, with a further flattening curve towards the end of the sample.\newline

\begin{table}[]
\centering
% \label{features}
\begin{tabular}{l|c|c|c|c|c|c}
%\toprule
\hline
Date & 03-09 & 03-16 & 04-12 & 04-19 & 05-21 & 05-28  \\
\hline
%\midrule
Infected Patients    &     1187  & 4907  & 54110  & 70715  & 167046 & 183746    \\
Treated Patients     &      7985 &   23073  & 102253  & 108257  & 60960   & 47986     \\
RSFE Treated   &      226   & 1156   & 1405   &  224  &  195 & 1958 \\
RSFE Confirmed &    292  &  1562  &  2619  &   199 &  18980  & 25316 \\
\hline
%\bottomrule
\end{tabular}
\caption{Selected data points for predicted number of infected and treated patients, as well as the MRSFE. Results are obtained using the hybrid assimilation covariance matrix for Italy}
\label{tab:table_hybridcov_it}
\end{table}{}

\begin{table}[]
\centering
% \label{features}
\begin{tabular}{l|c|c|c|c|c|c}
%\toprule
\hline
Date & 03-09 & 03-16 & 04-12 & 04-19 & 05-21 & 05-28  \\
\hline
%\midrule

% Infected Patients    &     21 &    85 &     12629 &    18492 &    36042 &    37837     \\
% Treated Patients     &      300 &     1458 &     71650 &     101575 &    214866 &    231290     \\
% RSFE Treated &       261 &  110 & 2049 & 1592 & 1308 &  498 \\
% %RSFE Infections & 4.30000e+01 & 1.14000e+02 & 4.43900e+03 & 1.39500e+04 & 1.33162e+05 & 1.72417e+05 \\
% RSFE Infections & 43 & 114 & 4439 & 1395 & 133162& 172417 \\

Infected Patients    &     31 &    154 &     16755 &    32023 &    168831 &    209999   \\
Treated Patients     &      290 &     1405 &    72363 &     101051 &    214629 &    232074     \\
RSFE Treated &       10&  53 & 713 & 524 & 237 &  784 \\
RSFE Confirmed & 10 & 69& 4126 & 1353 & 132789& 172162 \\

\hline
%\bottomrule
\end{tabular}
\caption{Selected data points for predicted number of infected and treated patients, as well as the MRSFE. Results are obtained using the hybrid assimilation covariance matrix for the UK}
\label{tab:table_hybridcov_uk}
\end{table}{}

\begin{table}[]
\centering
% \label{features}
\begin{tabular}{l|c|c|c|c|c|c}
%\toprule
\hline
Date & 03-09 & 03-16 & 04-12 & 04-19 & 05-21 & 05-28  \\
\hline
%\midrule
Infected Patients    &     29 &  117  & 59074    & 111282  &    393120   & 501607    \\
Treated Patients     &      559  &  4544   & 496239   & 647527   & 1184027   & 1220146     \\
RSFE Treated   &       32   &   152  &  7806 & 3911 & 13667  & 15347 \\
%RSFE Infections &   3.10000e+01  & 2.81000e+02  & 6.30550e+04  & 1.13096e+05  & 6.16649e+05  &  7.26317e+05 \\
RSFE Confirmed &   31 & 281  & 630550 & 113096  & 616649 &  726317 \\

\hline
%\bottomrule
\end{tabular}
\caption{Selected data points for predicted number of infected and treated patients, as well as the MRSFE. Results are obtained using the hybrid assimilation covariance matrix for the US}
\label{tab:table_hybridcov_us}
\end{table}{}

This is in contrasts to the United States and the United Kingdom as is visible in Fig.~\ref{fig:SITR_us} and Fig.~\ref{fig:SITR_uk} where the number of latent infections has been relatively constant and exhibits less of an downward trending behaviour. With the number of hospitalized infections increasing, this is likely due to the early relaxation and less stringent quarantine restriction in the United States compared to Italy. When inspecting both infected and hospitalized compartments, results of the SITR model aligns with policy choices, with a flattening tendency for Italy, contrasting with results for the United States and the United Kingdom which followed later or with less rigorous quarantine restrictions. Table \ref{tab:table_hybridcov_uk} depicts selected entries and forecasting errors for the United Kingdom, where at the end of the sample in May the amount of latent infections is estimated to be around 209,000.

Table \ref{tab:table_hybridcov_us} shows selected values for the United States. Even taking account the higher amount of infection numbers in the United States, the high amount of prediction errors is noticeable, indicating that the data is noisy ad the model fit is not performing as well as in the other two cases.
\newline 
Given these results the number of infections in the United States and the United Kingdom are likely to keep on growing, especially if the government is not considering the implementation of tighter regulations and quarantine measures. The results furthermore demonstrate how the assimilation framework can be extended to multiple countries and provide robust results given the large uncertainty in infection estimates.

\subsection{Lockdown Effects on Transmissibility}
To compare the predicted dynamics of our model estimates and to evaluate policies, we extend the analysis and discuss the dynamics of the model parameters. 
Following the same framework, we analyze the estimated transmissibility rate over time in all three countries for which we depict weekly averaged results. 
Italy has seen the earliest surge of Covid19 cases but kept its quarantine measures intact.
The US and UK were followed by rising surges with a delay, but have had a looser approach to quarantine measures.

\begin{figure}[!htb]
\centering
  \includegraphics[width=0.5\linewidth]{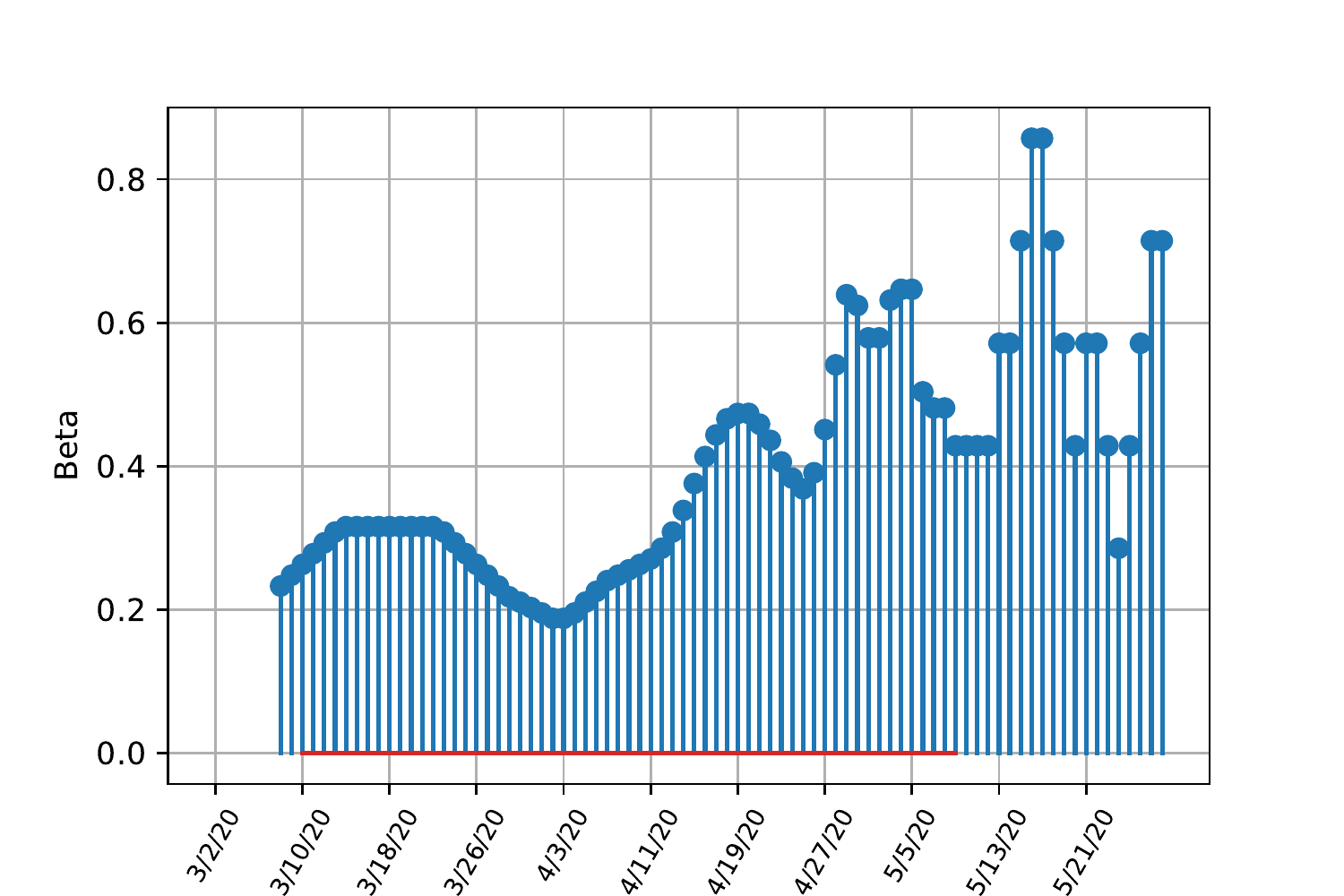}
  \caption{SITR short run transmissibility dynamics for Italy }\label{fig:it_short_beta}
\label{fig:SITR_it_beta}
\end{figure}

As given in Fig.~\ref{fig:it_short_beta}, the infection rate $\beta$ reached a high level of 0.3 from March 10th to to around March 20th, followed by a gradual decreasing period with an infection rate bottoming out at a value of 0.2, showing initial successes in lockdown measures, which were enforced from the 9th of March onwards. This shows how the lockdown order has first stabilized the transmissibility rate and later led to a decrease. Towards the end of the sample with increasing relaxation of the lockdown measures an increase in $\beta$ is observable. Variation of the parameters is very high towards the end of the sample, especially compared to examples of the UK and US, being evidence that the model parameter estimates are less clear on a strong increase in infection rates as in the US or the UK. 
%The predicted infected people in the community touched the peak on Feb 3rd with a daily incidence around 440. When comparing parameter estimates in Fig.~\ref{fig:us_short} to the estimates of the first example, the $\beta$ in the second example is smaller and the incidence peak is 3 days earlier, which indicated the effects of intervention in early epidemic stages. 
%Preliminary Results

\begin{figure}[!htb]
\centering
  \includegraphics[width=0.5\linewidth]{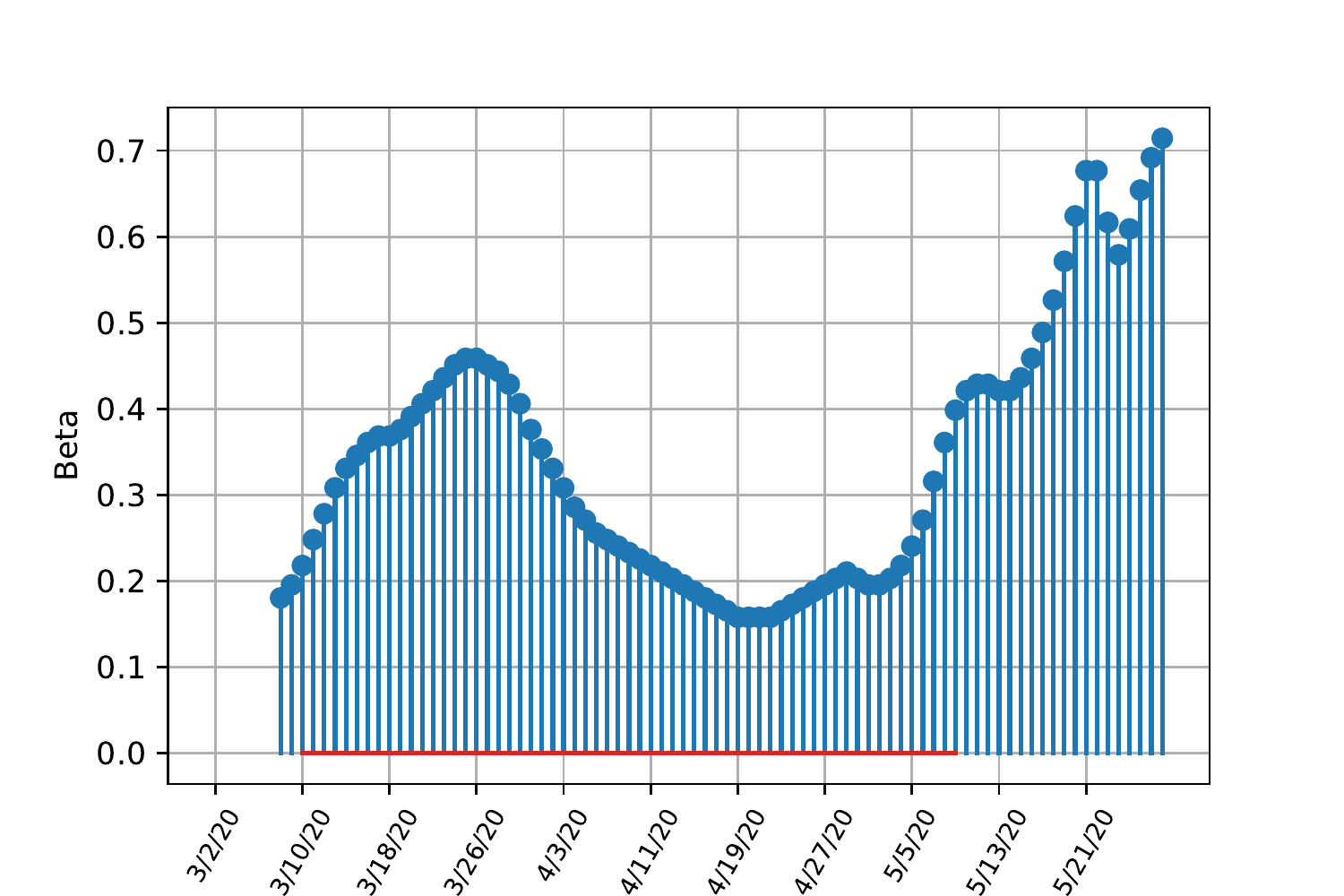}
  \caption{SITR short run transmissibility dynamics for the United States }\label{fig:us_short_beta}
\label{fig:SITR_us_beta}
\end{figure}

Overall,the Italian dynamics in Fig.~\ref{fig:it_short} and Fig.~\ref{fig:it_short_beta} clearly indicate that the number of undetected latent infections is decreasing, with also the number of known and hospitalized cases having crossed their peak value as well and approaching a very low value in the sample. The infection rates indicate a decrease after the enforcement of quarantine measures, with a increase in transmissibility towards the end of the sample, although the model parameters exhibit strong variation in parameter estimates.

The United States have not reached a similar level as Italy.
The trajectories in Fig.~\ref{fig:us_short_beta} illustrate that after a rapid growth and high peak value of transmissibility on the 26th of March, a strong decrease followed, showing the effectiveness of the lockdown measures that were enforced in many states from the 23th of March onwards.
After an initial strong decrease of the transmissibility rate after imposing a lockdown, the infection rate has strongly increased again at the end of the sample, with the transmissibility values increasing from a trough of 0.18 on the 19th of April to 0.7 at the end of May. This, together with the trajectories in Fig.~\ref{fig:us_short} show that restriction measures have only lead to initial successes with a stabilization of latent infection cases with the total number of cases still increasing. Especially towards the end of the sample at the end of May the amount of infections is increasing again. \newline

\begin{figure}[!htb]
\centering
  \includegraphics[width=0.48\linewidth]{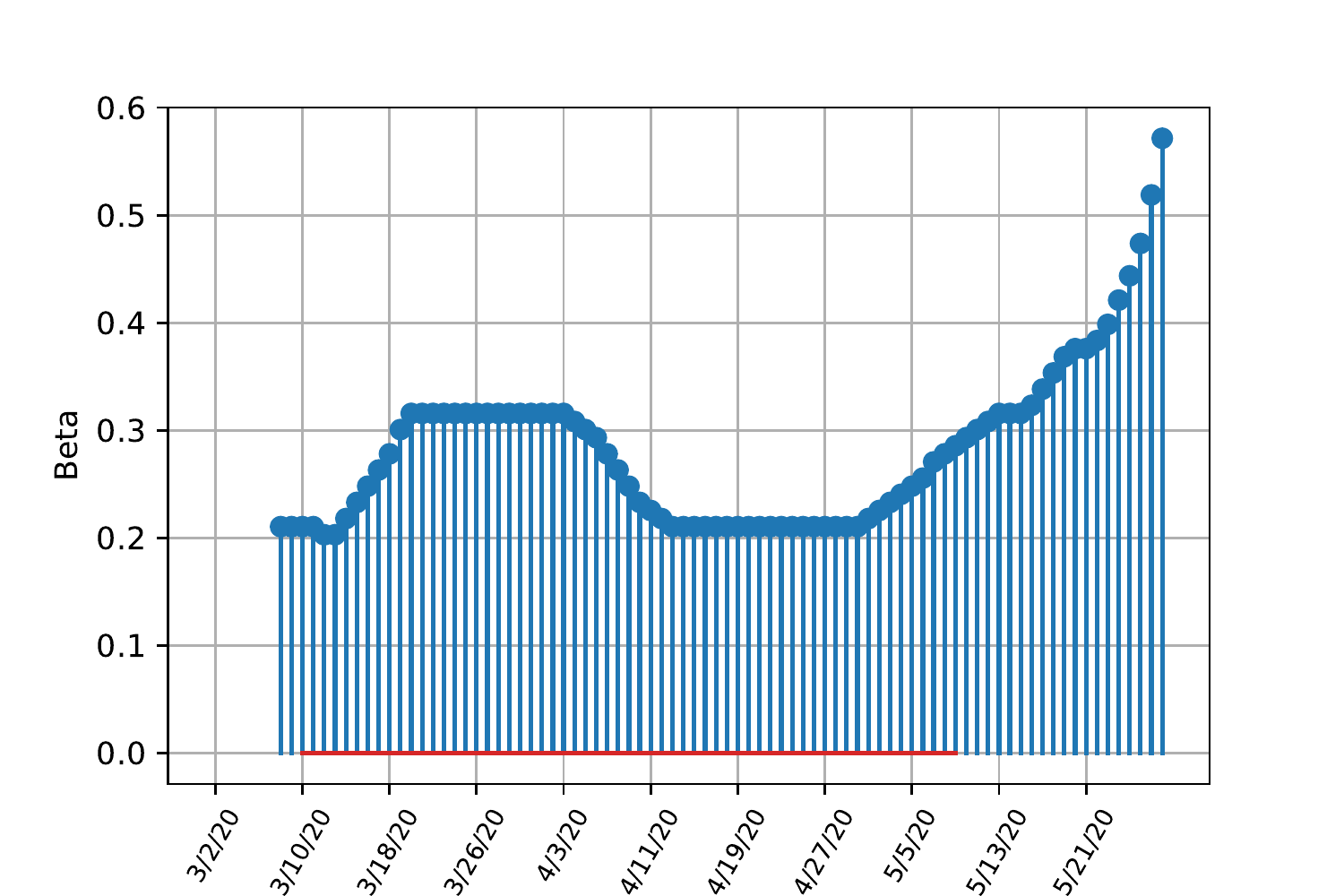}
  \caption{SITR short run transmissibility dynamics for the United Kingdom}\label{fig:uk_short_beta}
\label{fig:SITR_uk_beta}
\end{figure}

The development of the United Kingdom is very similar to the United States, where initial lockdown measures managed to decrease the number of latent infections, with the initial growth in infection numbers in the middle of March is not as strong as in the Italy or the US. This is explainable when taking into account the parameter estimates given in Fig.~\ref{fig:uk_short_beta}. 
The parameter plot visualises this development clearly, where transmissibility has initially decreased from the 4th of April onward and stabilized at a level of 0.2 and steadily increased from the end of April onwards at the end of the sample. The parameter estimates unequivocally indicate a strong increase in infection rates, with the last estimated peaking at around 0.6, although the relative increase from it's lowest value throughout April is not as pronounced as in the US indicating that the increase in transmissibility has not been affected as much by changing policies or behaviour in the population as in the US.\newline

Overall the results indicate that Italy is progressing well in containing the virus, the UK and especially the US are struggling to contain the virus in the medium-term. Especially the parameters indicate worsening developments, where in all three countries transmissibility has decrease initially due to quarantine measures but increased towards the end of the sample, although evidence is less clear and more uncertain for the Italian transmissibility rates.

%POLICY BLABLABLA, include a table? lockdownm uk start:23.3, italy start 9.3 , us start 23.3

%The long range forecasts in Fig.~\ref{fig:SITR_bj_long} indicate a gradual decrease of the number of treated patients as well as the number of infections, resembling the dynamics for Wuhan on a lower level.\newline Compared to Wuhan, the peak of the infection occurs five days before Wuhan on the 2nd of February, showing that low early infection levels as well as quarantine measures introduced by the government led to a rapid decline of infection cases. To put results into perspective, the next section applies the methodology to international data, giving estimates of peaks of covid19 globally.

\section{Conclusion and future work}
We introduced a novel epidemiological assimilation scheme to forecast and evaluate the current corona pandemic worldwide with a specific focus on the United States, the United Kingdom and Italy. We combined compartmental models in epidemiology with data assimilation schemes showing the advantage of real-time forecasting and parameter estimation in the current crisis. We discussed the benefits and differences in infection numbers when models are updated on a daily basis compared to static modelling. We then introduced a model extension allowing us to observe patients being treated, and patients being removed from the infectious population, which we labelled SITR.
Since models are sensitive to estimates of the covariance matrices, we add a hybrid ensemble approach which allows for robust covariance matrix estimates. We find that in Italy the peak of infections has been reached already, with the number of patients being treated peaking middle of April. The trajectories of the US and UK are less clear, with a likely increase in the medium term, with both countries showing a strong increase in transmissibility rates after an initial decrease due to lockdown measures. \newline
The generalisability of our model allows the addition of different compartments to the model, and also allows for the implementation in a variety of cases and countries, where in our experiments the model gives forecasts and parameter estimates for three different countries. Since this work focused mainly on the methodology of providing a robust recursive Bayesian estimation for the current nCov-2019 outbreak,  we propose a further in depth-study of the parameter estimates and an extended comparative study across countries.
Future work can add further complexities to the model, such as taking into account different mortality rates due to population age, cultural norms or quality of the healthcare system, providing applicability and robustness of the model for different datasets and scenarios.\newline
We encourage both researchers and policymakers to run similar test results with data from other countries or on a more local level to estimate potential infection rates of outbreaks and the rate of transmission to implement the correct policy measures to contain and mitigate adverse effects of the pandemic.

\appendix
%\begin{appendices}
\section{Appendix}

\subsection{Hybrid Data Assimilation}
In this section we describe the computation of the covariance matrices as explained in section 5. We estimate values for both the state and observation covariance matrices  $\textbf{Q}$ and  $\textbf{P}$ by using an ensemble approach \cite{wang2013gsi, lim2019hybrid}. The values for $\textbf{P}$ are based on an estimate of the residual covariance matrix of the stationary observed time series. Following the cost function give by Eq.~\ref{eq:3}, with $\textbf{x}^{b}$ representing an individual background state vector, $\textbf{x}^{b} = [S,I,T,R]$. 
The full ensemble of state vectors is given by
\begin{equation}
    \textbf{x}_{(1)}^{b},\textbf{x}_{(2)}^{b},..,\textbf{x}_{(N)}^{b}
\end{equation}
If the ensemble mean is defined as $\overline{\textbf{x}}^{b}$, then $\textbf{V}_{ens}$, the background state perturbations are computed via
\begin{equation}\label{eq:3.30}
    \textbf{V}_{ens} = \textbf{X}^{b}=\frac{1}{\sqrt{N-1}}(\textbf{x}_{(1)}^{b}-\overline{\textbf{x}}^{b},\textbf{x}_{(2)}^{b}-\overline{\textbf{x}}^{b},...,\textbf{x}_{(N)}^{b}-\overline{\textbf{x}}^{b})
\end{equation}
In this case, $\textbf{V}_{ens}$ and $\textbf{X}^{b}$ are a $n$ x $N$ matrix called the ensemble background perturbation matrix. The rank-deficient version of the background error covariance matrix is defined as $\textbf{Q}^{*}$ with
\begin{equation}\label{eq:3.30}
    \textbf{Q}^{*} = {\textbf{X}^{b}}^{T}\textbf{X}^{b}
\end{equation}

The ensemble is static, meaning that it does not evolve dynamically with time, but it still incorporates flow-dependent information at the start time which is still beneficial for an extended Kalman filter or 4D analysis.% which is assuming no time dependence.

The way the ensembles are chosen and computed determines the accuracy of ensemble DA.

The ensemble needs to be computed in such a way that the time dependent variability of the background error covariance matrix, as well as the correlation of variables is captured by the sampling procedure.

The method we devise is to divide the collection of background states, $\underbar{\textbf{x}}^{b}$ based on the size of the ensemble into $N$ equally sized groups with each group being denoted by $\underbar{\textbf{x}}^{b}_{(i)}$ meaning that ensemble members belong to the $i$th group. The mean and standard deviation of each group is then estimated and used to sample the ensemble members from.
\begin{algorithm}[H]
%\SetAlgoLined
 \caption{Build Ensemble}\label{algo3}
 \begin{algorithmic}[1]
 \STATE Inputs: $\underbar{\textbf{x}}^{b}$
\STATE $i=0$, $N = \text{ensemble size}$, $n = length(\textbf{x}^{b})$
\FOR {$\underbar{\textbf{x}}^{b}_{(i)}$ in $array\_split(\underbar{\textbf{x}}^{b}, N)$}
    \STATE $\mu_{(i)} =  mean(\underbar{\textbf{x}}^{b}_{(i)})$
    \STATE $\sigma_{(i)} = standard\_deviation(\underbar{\textbf{x}}^{b}_{(i)})$
    \STATE $ensemble[:,i] = normal\_distribution(\mu_{(i))},\sigma^{2}_{(i)})$
    \STATE $i = i+1$
\ENDFOR
\STATE $ensemble\_mean = mean(ensemble)$
\FOR{$i=0,1,..,N$}
\STATE $\textbf{V}_{ens}[:,i] = ensemble[:,i]-ensemble\_mean$
\ENDFOR
\RETURN $\textbf{V}_{ens}$
\end{algorithmic}
\end{algorithm}

Algorithm~\ref{algo3} describes in detail how $\textbf{V}_{ens}$  is computed and ensembles are formed. The full background state matrix, $\underbar{\textbf{x}}^{b}$ is split into $N$ groups each of size $n\times \frac{n}{N}$. Both, the  means as well as the standard deviations of the $n$ rows are estimated and used to generate draws from a multivariate Gaussian distribution to form the ensemble. In order to form $\textbf{V}_{ens}$, for each ensemble member the corresponding mean is estimated and then subtracted, computing the standard deviation.
To put results into perspective we discuss the difference between standard assimilation and hybrid approaches by conducting a sensitivity analysis next.

\section*{Acknowledgements}
We are grateful for helpful discussions and feedback from Joseph Wu, Neil Ferguson and other participants at the Royal Society based DSI workshop "Scientists against CoViD-19 and beyond"%\vspace*{-12pt}
\section*{Version of Record}
The initial conference version was accepted at the International Conference of Machine Learning (ICML) HSYS 2020 workshop. This extended journal version was published in the European Journal of Epidemiology. Please cite as: Nadler, P., Wang, S., Arcucci, R. et al. An epidemiological modelling approach for COVID-19 via data assimilation. Eur J Epidemiol 35, 749–761 (2020). https://doi.org/10.1007/s10654-020-00676-7

%\end{appendices}

\bibliographystyle{unsrt}  
\bibliography{references}  %%% Remove comment to use the external .bib file (using bibtex).

\begin{thebibliography}{10}

\bibitem{asch_2016}
Mark Asch, Marc Bocquet, and Maelle Nodet.
\newblock {\em Data assimilation: methods, algorithms, and applications}.
\newblock 12 2016.

\bibitem{imai2020estimating}
Natsuko Imai, Ilaria Dorigatti, Anne Cori, Steven Riley, and Neil~M Ferguson.
\newblock Estimating the potential total number of novel coronavirus cases in
  wuhan city, china, 2020.

\bibitem{li2020early}
Qun Li, Xuhua Guan, Peng Wu, Xiaoye Wang, Lei Zhou, Yeqing Tong, Ruiqi Ren,
  Kathy~SM Leung, Eric~HY Lau, Jessica~Y Wong, et~al.
\newblock Early transmission dynamics in wuhan, china, of novel
  coronavirus--infected pneumonia.
\newblock {\em New England Journal of Medicine}, 2020.

\bibitem{wu2020nowcasting}
Joseph~T Wu, Kathy Leung, and Gabriel~M Leung.
\newblock Nowcasting and forecasting the potential domestic and international
  spread of the 2019-ncov outbreak originating in wuhan, china: a modelling
  study.
\newblock {\em The Lancet}, 2020.

\bibitem{rhodes2009variational}
CJ~Rhodes and T~D{\'e}irdre Hollingsworth.
\newblock Variational data assimilation with epidemic models.
\newblock {\em Journal of theoretical biology}, 258(4):591--602, 2009.

\bibitem{bettencourt2007towards}
Luis~MA Bettencourt, Ruy~M Ribeiro, Gerardo Chowell, Timothy Lant, and Carlos
  Castillo-Chavez.
\newblock Towards real time epidemiology: data assimilation, modeling and
  anomaly detection of health surveillance data streams.
\newblock {\em NSF Workshop on Intelligence and Security Informatics}, pages
  79--90, 2007.

\bibitem{wang2013gsi}
Xuguang Wang, David Parrish, Daryl Kleist, and Jeffrey Whitaker.
\newblock Gsi 3dvar-based ensemble--variational hybrid data assimilation for
  ncep global forecast system: Single-resolution experiments.
\newblock {\em Monthly Weather Review}, 141(11):4098--4117, 2013.

\bibitem{bonavita2016evolution}
Massimo Bonavita, Elias H{\'o}lm, Lars Isaksen, and Mike Fisher.
\newblock The evolution of the ecmwf hybrid data assimilation system.
\newblock {\em Quarterly Journal of the Royal Meteorological Society},
  142(694):287--303, 2016.

\bibitem{bettencourt2008real}
Luis~MA Bettencourt and Ruy~M Ribeiro.
\newblock Real time bayesian estimation of the epidemic potential of emerging
  infectious diseases.
\newblock {\em PLoS One}, 3(5), 2008.

\bibitem{cobb2014bayesian}
Loren Cobb, Ashok Krishnamurthy, Jan Mandel, and Jonathan~D Beezley.
\newblock Bayesian tracking of emerging epidemics using ensemble optimal
  statistical interpolation.
\newblock {\em Spatial and spatio-temporal epidemiology}, 10:39--48, 2014.

\bibitem{li1995global}
Michael~Y Li and James~S Muldowney.
\newblock Global stability for the seir model in epidemiology.
\newblock {\em Mathematical biosciences}, 125(2):155--164, 1995.

\bibitem{seibert2019validation}
Jan Seibert, Maria Staudinger, and HJ~Ilja van Meerveld.
\newblock Validation and over-parameterization—experiences from hydrological
  modeling.
\newblock In {\em Computer Simulation Validation}, pages 811--834. Springer,
  2019.

\bibitem{anderson1991discussion}
Roy~M Anderson.
\newblock Discussion: the kermack-mckendrick epidemic threshold theorem.
\newblock {\em Bulletin of mathematical biology}, 53(1-2):1, 1991.

\bibitem{cuomo2017numerical}
Salvatore Cuomo, Ardelio Galletti, Giulio Giunta, and Livia Marcellino.
\newblock Numerical effects of the gaussian recursive filters in solving linear
  systems in the 3dvar case study.
\newblock {\em Numerical Mathematics: Theory, Methods and Applications},
  10(3):520--540, 2017.

\bibitem{nadler2019scalable}
Philip Nadler, Rossella Arcucci, and Yi-Ke Guo.
\newblock A scalable approach to econometric inference.
\newblock In {\em PARCO}, pages 59--68, 2019.

\bibitem{arcucci2017variational}
Rossella Arcucci, Luisa D'Amore, Jenny Pistoia, Ralf Toumi, and Almerico Murli.
\newblock On the variational data assimilation problem solving and sensitivity
  analysis.
\newblock {\em Journal of Computational Physics}, 335:311--326, 2017.

\bibitem{JHUdata}
John hopkins university coronavirus resource center.
\newblock \url{https://coronavirus.jhu.edu/map.html}, 2020.
\newblock Accessed: 2020-05-24.

\bibitem{CDCdata}
National health commission of the people’s republic of china.
\newblock \url{http://web.archive.org/web/20080207010024/}.
\newblock Accessed: 2020-02-07.

\bibitem{lim2019hybrid}
Edward~M Lim, Miguel~Molina Solana, Christopher Pain, Yi-Ke Guo, and Rossella
  Arcucci.
\newblock Hybrid data assimilation: An ensemble-variational approach.
\newblock In {\em 2019 15th International Conference on Signal-Image Technology
  \& Internet-Based Systems (SITIS)}, pages 633--640. IEEE, 2019.

\end{thebibliography}

\end{document}